\documentclass[letterpaper]{JHEP3}
\usepackage{amsmath}
\usepackage{amssymb}
\usepackage{amsfonts}
\usepackage{lscape, graphicx}

\newcommand{\Dslash}{\,{\raise.15ex\hbox{/}\mkern-12mu D}}

\renewcommand{\bar}{\overline}
\newcommand{\tr}{\textrm{tr}}

\newcommand{\bea}{\begin{eqnarray}}
\newcommand{\eea}{\end{eqnarray}}
\newcommand{\be}{\begin{equation}}
\newcommand{\ee}{\end{equation}}
\newcommand{\iless}[1]{\bigg\lfloor #1 \bigg\rfloor}

\title{Index theorem for  topological excitations on $\mathbf{R^3 \times S^1}$ 
and Chern-Simons theory }

\author
    {
    {
    \def\href#1#2{#2}	
    Erich Poppitz$^1$\footnote{\email{poppitz@physics.utoronto.ca}}~
    and Mithat \"Unsal$^2$\footnote{\email{unsal@slac.stanford.edu}}~
           \\${}^1$Department of Physics, University of Toronto,
    Toronto, ON M5S 1A7, Canada
     \\${}^2$SLAC and Physics Department, Stanford University, Stanford, CA 94025/94305, USA
        }
    }%

\abstract{

\small
We derive an  index theorem for  the Dirac operator in the background of 
various topological excitations  on an $R^3 \times S^1$ geometry. 
   The index theorem provides 
  more refined data than the APS index  for an instanton on $R^4$ and reproduces it in decompactification limit. In the $R^3$ 
 limit, it reduces to the Callias index theorem.   
 The index is expressed in terms of topological charge and the $\eta$-invariant associated with the boundary Dirac operator. Neither topological charge  nor $\eta$-invariant 
  is typically an integer, however,  the non-integer parts cancel to give an integer-valued index. 
Our derivation is based 
on axial current non-conservation---an exact operator identity valid on any four-manifold---and  on the existence of a   center symmetric, or approximately center symmetric, boundary  holonomy (Wilson line).  
We expect the index theorem  to usefully   apply to many physical systems of interest, such as   low temperature (large $S^1$, confined) phases of gauge theories,  
center stabilized  Yang-Mills theories with vector-like or chiral matter (at  $S^1$ of any size), and 
 supersymmetric gauge theories with supersymmetry-preserving boundary conditions (also at any $S^1$). 
In QCD-like and chiral gauge theories, the  index theorem should shed light into the nature of topological 
excitations responsible  for chiral symmetry breaking and the generation of mass gap in the gauge sector. 
We also show that imposing chirally-twisted  boundary condition in gauge theories with fermions 
induces a Chern-Simons term in the infrared.  This suggests that  some QCD-like gauge theories 
should possess components with  a topological Chern-Simons phase in the small $S^1$ regime.  
      }

\begin{document}

\maketitle

\section{Introduction}

\subsection{Motivation}

 A method to study non-perturbative aspects of an asymptotically free gauge theory on $R^4$ 
 is to begin  with a compactification on  $R^3 \times S^1$.  
  $S^1$ may be either a spatial   or temporal circle, determined according to the spin connection of fermions.  
In  asymptotically free gauge theories, the size of $S^1$  is  a control parameter for the 
strength of the running coupling at the scale of compactification.  At radius much smaller than the strong  length scale, the theory is weakly coupled and at large radius, it is strongly coupled. 
It is well-known that certain aspects of   weakly coupled gauge theories are amenable to perturbative treatment. It is  less known  that such gauge theories  also admit a 
semi-classical non-perturbative treatment if the boundary Wilson line (Polyakov loop) satisfies 
certain conditions. 

Let $U( x)$ denote the holonomy of Wilson line wrapping the $S^1$ circle: 
\be
U( x) = P\exp{ i \oint A_4 dy}, \qquad x \in R^3, \; \; y \in S^1~,
\ee 
where $A_4( x, y)$ is the component of the gauge field along the compactified direction.  If the eigenvalues of the $A_4$ field (which are gauge invariant) repel each other in the weak coupling regime, for a dynamical reason or due to a deformation  described below,  
then the  boundary value of the $A_4$ field  as $|x| \rightarrow \infty$ takes the form:
\be
\label{A4inf}
A_4\big\vert_\infty =  {\rm diag}(\hat{v}_1, \hat{v}_2 , \ldots \hat{v}_N)~,  \qquad 
 \hat{v}_1 < \hat{v}_2 <\ldots < \hat{v}_N ~.
\ee
In the weak coupling regime, $A_4$  behaves as a compact adjoint Higgs field and 
the vacuum configuration (\ref{A4inf}) induces  gauge symmetry breaking, 
$G \rightarrow {\bf Ab}(G)$, to the maximal abelian subgroup  ${\bf Ab}(G)$.  This means that there  exist
a plethora of  stable topological excitations in such four dimensional gauge theories, such as magnetic monopoles, magnetic bions, instantons  and other interesting (stable) composites. 

It is well-known that in a thermal set-up at sufficiently high temperatures (and weak coupling), dynamics disfavors configurations for Wilson lines such as (\ref{A4inf}) \cite{Gross:1980br}, and favors configurations for which $A_4\big\vert_\infty =  {\rm diag}(\hat{v}, \hat{v} , \ldots \hat{v})~= 
(0,0, \ldots,0)$. 
In such cases, the effect of topological excitations is suppressed  by volume factors, and 
 semi-classical techniques do not usefully apply    \cite{Gross:1980br}.
  Due to these legitimate  reasons, semi-classical methods in finite temperature setting have not received   wide attention so far (although see \cite{Diakonov:2007nv}), and 
 are not part of the common-place techniques to study  Yang-Mills theory and 
  non-supersymmetric Yang-Mills theory with vector-like and chiral matter, 
   compactified on a circle. 
 
 However, there are at least three ways to make such boundary values of Wilson lines stable  at weak coupling. These are:
 {\it a}.) center-stabilizing double-trace deformations, 
{\it b}.) adjoint fermions with periodic boundary conditions, or mixed representations of adjoints and  a  few complex representation fermions all with periodic boundary conditions, and 
{\it c}.) supersymmetry and supersymmetry preserving boundary conditions.
In this sense, the case of non-trivial holonomy (\ref{A4inf}) at weak coupling 
 is as generic as the high-temperature trivial  holonomy.  
In particular, with the center stabilizing double-trace deformations, certain   gauge 
theories on $R^4$, such as Yang-Mills theory and vector-like and even chiral theories can be smoothly connected to small $S^1 \times R^3 $ \cite{Shifman:2008ja, Unsal:2008ch, Shifman:2008cx}.  There already exist evidence from lattice gauge theory  (where deformations were also suggested independently to explore phases of partial center symmetry breaking)  
that the  conjecture of smoothness holds  for Yang-Mills theory 
\cite{Myers:2007vc}.\footnote{The center-stabilized 
small-$S^1$ regime of gauge theories  is amenable to both numerical lattice simulations  and 
non-perturbative semi-classical techniques. In this sense, this regime provides a first example in which we can confront a controlled approximation, including non-perturbative effects,  with the lattice, and is, in our opinion, an important opportunity for both lattice and continuum gauge field theory.} 
Therefore, there is currently a strong incentive to study in detail the topological excitations on  
$R^3 \times S^1$ and the index theorems associated with these excitations.

Our interest is in the index of the Dirac (or Dirac-Weyl) 
 operator  $\hat{D}$ (\ref{vdirac})  in the background of 
topological excitations pertinent to the gauge theory on $R^3 \times S^1$.  The reason 
that this is interesting for non-perturbative physics is two-fold: 
\begin{itemize}
\item{ The topological excitations  
with non-vanishing index will carry compulsory fermion zero modes attached to them, and may induce chiral symmetry breaking. Generically,  the fermionic index of a monopole operator on 
$R^3 \times S^1$ is (much)  smaller than the APS index for the BPST instanton.  
 Thus, they are in principle more relevant for low energy phenomena.}
\item{The generation of  mass gap (and confinement) for gauge fluctuations, in the weak coupling regime,  requires the existence of topological excitations with  vanishing index. In typical  QCD-like and chiral gauge theories, most of the leading topological excitations (monopoles)  carry fermionic zero modes, hence cannot contribute to the mass gap. Therefore, the index theorem can be used to identify composite topological excitations (such as magnetic bions) for which the sum of individual indices add up to zero. }
\end{itemize}
A  simple example which illustrates both issues is Yang-Mills theory with adjoint fermions (QCD(adj)) and   ${\cal N}=1$  SYM. 
In both cases, magnetic monopole operators (which appear at order $e^{-S_0}= e^{-8 \pi^2/ (g^2N)}$  in the semi-classical  $e^{-S_0}$ expansion)   induce a certain chiral condensate. However, the topological excitations responsible for the existence of mass gap of the dual photon and thus for confinement are the magnetic bions  with vanishing 
index, which appear at order $e^{-2S_0}$.   The index theorem on $R^3 \times S^1$ should help us identify both classes of non-perturbative 
topological excitations for any gauge theory.

 The non-perturbative semi-classical analysis   provides reliable information about the gauge theory in the weak coupling regime.  However,  the semi-classical treatment does not extend over to the large radius, 
strong coupling regime, where  the eigenvalues of $A_4$ fluctuate rapidly, and there is no ``Higgs regime" where the long-distance theory abelianizes.  In the partition function, we need to sum over all gauge inequivalent configurations. 
At  $x=\infty$,  $A_4$ field can acquire a profile consistent with the unbroken center symmetry, such as   (\ref{A4inf}).   In fact,  the  boundary Wilson line 
 (\ref{A4inf})   defines an isotropy group at infinity \cite{Gross:1980br}: 
 $G_{A_4|_{\infty}}=  \{ g  \in G \; | \;  gU(\infty)g^{\dagger}= U(\infty)\}$. For example, in low temperature pure  Yang-Mills theory, 
 the isometry group is isomorphic to the maximal abelian subgroup, 
 $G_{A_4|_{\infty}} \sim {\bf Ab}(G)$. This does not mean that 
 a dynamical abelianization takes place in this regime, nor semi-classical techniques apply. 
 However,  the index theorem for the Dirac operator can be interpolated from $R^3$ to $R^4$. The index theorem is valid at any radius, regardless of the value of the coupling constant. 
 
  Although the index theorem and topological excitations consistent with the isotropy group 
  $G_{A_4|_{\infty}} $ continue to exist in the large $S^1$ strong-coupling domain,  the semi-classical techniques no longer usefully apply. 
   Nonetheless, we believe that  there is  value in studying the  form of the 
  topological operators, dictated by the appropriate index theorem, and at least qualitatively study  their dynamical effects.  This is the goal of the 
  recent  ``deformation program." 
 
 \subsection{Outline}

We introduce our notation in   Section 1.3 below. 
 In Section 2, we begin the calculation of the index. Our calculation can be thought of as a generalization of that of \cite{Weinberg:1979ma, Weinberg:1979zt}, see also \cite{Niemi:1985ht}.
We show that the index on $R^3 \times S^1$ has two contributions---a topological charge and surface term contribution. 

In Section 2.1, we first calculate the index for static monopole backgrounds.  The surface term contribution, Section 2.1.1, is expressed in terms of the $\eta$-invariant of the boundary Dirac operator, while the topological charge contribution is given in Section 2.1.2. The final formula for the index in the ``static" background  is  eqn.~(\ref{indexstatic123}) for the fundamental of $SU(N)$  and  eqns.~(\ref{C1}--\ref{C4}) from   Appendix B for other representations.
The calculation of the index in a Kaluza-Klein (``winding") monopole background is given in Section 2.2, with the result for the fundamental of $SU(N)$   in (\ref{indexKKfund}), and in (\ref{C5}) for general representations.

 We note that an expression for the index on $R^3 \times S^1$ similar to ours---given in terms of the topological charge and the $\eta$-invariant---can be extracted from the appendix of ref.~\cite{Nye:2000eg}.  The contribution of this paper consists of: {\it a}.) a derivation of the index accessible to physicists along the lines  given  in the physics literature  for $R^3$ and 
  by using exact operator identities valid on any four-manifold and
  {\it b}.) a calculation of the index in specific backgrounds and a discussion of its jumps---properties which are   of interest for concrete quantum field theory applications.

In Section 3, we discuss in some more detail the index for the three lowest representations of $SU(2)$ and the fundamental of $SU(N)$. We explain  the jumps of the index which occur as the ratio of boundary holonomy to the size of $S^1$ is varied. 

In Section 4, we explain the relation to the Callias  \cite{Callias:1977kg} and APS \cite{Atiyah:1975jf} indices.

In Section 5, we consider  the generation of fermion-loop induced Chern-Simons terms on $R^3 \times S^1$. We show when Chern-Simons terms are induced  and how their coefficients are quantized. We consider the effect of   turning on of discrete Wilson lines for background fields gauging anomalous flavor symmetries (similar effects are known from the string literature). The resulting Chern-Simons terms have a profound effect on the phase structure of the theory on $R^3 \times S^1$.

Finally, Appendix A contains another calculation of the $\eta$-invariant. In Appendix B, we give formulae for the index for general representations.

\subsection{Notation}
\label{notation}

We take the four-dimensional Euclidean Dirac operator of a vector-like fermion in the representation ${\cal{R}}$ to be:
\be
\label{dirac1}
\hat{D} \equiv \gamma_\mu D_\mu, ~ D_\mu \equiv \partial_\mu + i A_\mu^a T^a~.
\ee
We use hermitean $T^a$'s, obeying Tr $T^a T^b = T({\cal{R}}) \delta^{ab}$, taking T(fund.)$=1/2$ for $SU(N)$. 
To further set and check notation, note that we use, in a given representation, $\psi \rightarrow U \psi$, $A_\mu \rightarrow U A_\mu U^\dagger - i U \partial_\mu U^\dagger$ under gauge transformations, hence
$F_{\mu\nu} = \partial_\mu A_\nu - \partial_\nu A_\mu + i [ A_\mu , A_\nu ]$. Roman indices  run from $1\ldots 3$, while Greek indices span $1 \ldots 4$; the $x^4 \equiv y$ direction is periodic, $y \equiv y+L$. 
The hermitean Euclidean $\gamma$-matrix basis we use is:
\be
\label{gammas}
\gamma_k = \sigma_1 \otimes \sigma_k~,~~ \gamma_4 = - \sigma_2 \otimes \sigma_0~, ~~ \gamma_5 = \sigma_3 \otimes \sigma_0~,
\ee where $\sigma_k$ are Pauli matrices and $\sigma_0$ is the unit matrix. The vector-like Dirac operator (\ref{dirac1}) is: 
\be
\label{vdirac}
\hat{D} = \left( \begin{array}{cc} 0 & \sigma_k D_k + i \sigma_0 D_4 \cr 
\sigma_k D_k - i \sigma_0 D_4 & 0 \end{array} \right)~ \equiv  \left( \begin{array}{cc} 0 & - D^\dagger \cr 
D & 0 \end{array} \right)~.
\ee
In the above equation, we defined the $2$$\times$$2$  Weyl operator  $D$, obeying:
\be
\label{DdaggerD}
D^\dagger D = - D_\mu D_\mu + \sigma^m {1\over 2} \epsilon_{mkl} F_{kl} + \sigma_k F_{4 k} 
= - D_\mu D_\mu + 2 \sigma^m B^m 
\ee
\be
D D^\dagger = - D_\mu D_\mu  + \sigma^m {1\over 2} \epsilon_{mkl} F_{kl} - \sigma_k F_{4 k} = - D_\mu D_\mu~.\label{DDdagger}
\ee
Here, $D_\mu$ is as defined in (\ref{dirac1}), and we assumed, without loss of generality, that the background of interest is anti self-dual, namely that $F_{4k} = {1\over 2} \epsilon_{kpq} F_{pq} \equiv B_k$ (all expressions can be easily generalized  for  self-dual backgrounds). In this paper, we will use ``Tr" to denote  traces of operators over spacetime as well as spinor indices, while ``tr" will refer to traces over spinor indices only.

We are interested in computing the index of the Dirac operator in topologically nontrivial backgrounds on $R^3 \times S^1$, generalizing the $R^3$ result  of \cite{Callias:1977kg}.
The simplest example of a nontrivial background is given by the three-dimensional $SU(2)$  Prasad-Sommerfield (PS) solution of unit magnetic charge, embedded in $R^3 \times S^1$. The other backgrounds on interest can be constructed by taking superpositions of the fundamental monopoles and other solutions, obtained by non-periodic ``gauge" transformations. 

The PS solution is  ``static" (i.e. $y$-independent) and the $A_4$-component of the gauge field plays the role of the Higgs field. 
For example, consider the anti self-dual solution, which 
 obeys $F_{4k} = - {1 \over 2} \epsilon_{4 kpq} F_{pq} = {1\over 2} \epsilon_{kpq} F_{pq} \equiv B_k$. In our conventions and in regular (``hedgehog") gauge, the $SU(2)$ solution reads:
\be
\label{PSsolution}
A_4 = A_4^a(r,v) \; T^a = \hat{r}^a  f(r, v)\; T^a~,~~ A_j = A_j^a(r,v)\; T^a = \epsilon_{jba} \hat{r}^b   g(r, v) \;T^a~,
\ee
where $\hat{r}^a = {r^a \over r}$ and:
\be
\label{PSsolution2}
f(r, v) = {1\over r} - v \coth v r \; , ~~ g(r, v) = - {1 \over r} + {v \over \sinh v r}~.
\ee
The asymptotics of the $B_k$, $A_4$ fields of the PS solution at infinity are:
\bea
\label{PSsolution3}
A_4\big\vert_{\infty} &=&  - v \; \hat{r}^a T^a \left(1 - {1\over v r} +\ldots\right)\nonumber\\
B_k\big\vert_{\infty} &=& {\hat{r}^k \over r^2}\; \hat{r}^a T^a + \ldots  , 
\eea
where dots denote terms that vanish as $e^{- v r}$. To cast the solution in string gauge, we need to gauge transform $\hat{r}^a T^a \rightarrow T^3$. The asymptotics of $A_4$ and $B_k$  in string gauge are obtained from (\ref{PSsolution3})  by replacing $\hat{r}^a T^a$ with $T^3$, for example  the asymptotics of  $SU(2)$-holonomy  is  $A_4\vert_\infty = {1\over 2} {\rm diag} (-v, v)$.

For the applications we have in mind, we want to also consider  monopole solutions of $SU(N)$. The $SU(2)$ PS solution considered above can be embedded in $SU(N)$ as described in \cite{Weinberg:1979zt}. For simplicity, we will use the fundamental generators of $SU(N)$ to describe the embedding (a description of the embedding using  roots and weights can be also given, see \cite{Weinberg:1979zt}; however, we find that for our purposes using $N\times N$ matrices is both sufficient and illuminating).
The general form of the asymptotics of the Higgs field $A_4$ is: 
\begin{eqnarray}
\label{A4infinity}
A_4\big\vert_\infty &=&  {\rm diag}(\hat{v}_1, \hat{v}_2 , \ldots \hat{v}_N)~, ~~ 
\nonumber  \\
&& \hat{v}_1 < \hat{v}_2 <  \ldots < \hat{v}_N , \qquad \sum_{i=1}^{N}  \hat{v}_i=0
\end{eqnarray}
where, without loss of generality, we have ordered the eigenvalues as in our $SU(2)$ example (for example, eqn.~(\ref{PSsolution3}) corresponds to taking 
 $\hat{v}_1$=$- \hat{v}_2$=$-{v\over 2}$).
A background with an additional overall $U(1)$ ``Wilson line" $a_0$, often also called ``real mass" term (when fermions are included),  allows
the holonomies $\hat{v}_j$ to be more general:
\be
\label{hatv1}\hat{v}_j  \rightarrow  \hat{v}_j    + {1 \over \sqrt {2N}}\; a_0 , 
\ee
where we also normalized the overall $U(1)$ generator multiplying $a_0$ to  Tr $T^2 = 1/2$. Including a non-vanishing $a_0$ can  be used to incorporate different   boundary conditions for the fermions in all our formulae.

The asymptotic form of the $U(N)$ holonomy  (\ref{A4infinity}, \ref{hatv1}) admits $N$ types of elementary monopoles; $N$-$1$ of these are associated with the positive simple roots ${\bf \alpha}_i$ of the $SU(N)$ Lie algebra,  for which:
\begin{eqnarray}
    {\bf \alpha}_i \cdot {\bf H} = \frac{1}{2} {\rm diag} (0, \ldots , \underbrace{1}_{i}, \underbrace{-1}_{i+1}, \ldots,0),  \qquad  i=1, \dots N-1~,
\label{eq:roots}
\end{eqnarray}
where ${\bf H}=(H^1, \ldots, H^{N-1})$, where $H^a$ denote the Cartan generators of $SU(N)$. The Cartan generators and simple roots are normalized as Tr $H^a H^b = {1 \over 2} \delta^{ab}$, $\alpha_i \cdot \alpha_j= \delta_{i , j} -{1 \over 2} \delta_{i, j \pm 1}$.
The $N^{\rm th}$ type of fundamental monopole  arises due to compactness of the ``Higgs" field  $A_4$, and    is associated with the ``affine"  root:
\begin{equation}
\label{affineroot}
    {\bf \alpha}_N \cdot {\bf H} \equiv  - \sum_{j=1}^{N-1} \, {\bf \alpha}_j \cdot {\bf H}~
    =  \frac{1}{2} {\rm diag} (-1, 0, 0, \ldots ,1) \,.
\end{equation}

A monopole solution corresponding to the $i^{\rm th}$ simple root (\ref{eq:roots}) of  $SU(N)$ can be constructed from (\ref{PSsolution3}) 
as follows. First, rewrite the holonomy (\ref{A4infinity}):
\be
\label{a4ith}
A_4\big\vert_\infty = {\rm diag}(\hat{v}_1, \hat{v}_2 ,\ldots, \hat{v}_{i-1}, \underbrace{V}_{i}, \underbrace{V}_{i+1}, \hat{v}_{i+2} ,\ldots , \hat{v}_N) + {1 \over 2} {\rm diag}(0, 0, \ldots,0,\underbrace{- \tilde{V}}_{i}, \underbrace{ \tilde{V}}_{i+1},0,\ldots , 0)~,
\ee
where $ V  = {1 \over 2} (\hat{v}_{i+1}+  \hat{v}_i )$ and $\tilde{V} =  \hat{v}_{i+1} - \hat{v}_{i}$. We now  diagonally embed the $SU(2)$ generators $\tau^a$ into $SU(N)$, such that their only nonzero elements are equal to one-half the Pauli matrices embedded in a $2 \times 2$ square along the diagonal of the $N\times N$ matrices (thus, their  diagonal elements are the $i$$^{\rm th}$ and $i$$+$$1$$^{\rm th}$ ones singled out in (\ref{a4ith})). With this embedding it is easy to explicitly verify that:
\begin{eqnarray}
\label{ithembedded}
A_4 &=& \hat{r}^a f(r, \tilde{V}) \tau^a  + {\rm diag}(\hat{v}_1, \hat{v}_2 ,\ldots, \hat{v}_{i-1}, \underbrace{V}_{i}, \underbrace{V}_{i+1} ,\hat{v}_{i+2},\ldots , \hat{v}_N)~,  \nonumber \\
A_m &=& \epsilon_{m b a} \hat{r}^b g(r, \tilde{V}) \tau^a ~,
\end{eqnarray}
with $f(r, \tilde{V})$ and $g(r, \tilde{V})$ defined in (\ref{PSsolution2}) 
solves the anti-self-duality condition $F_{4 k} = B_k$ inside $SU(N)$. The large-radius asymptotics can be immediately read off (\ref{PSsolution3}) by replacing $T^a$ with $\tau^a$ and inserting in (\ref{ithembedded}).

Finally,  a collection of $n_1, n_2,  ... , n_{N-1}$ fundamental monopoles of the type corresponding to the $1^{\rm st}$, $2^{\rm nd}$, ..., $N$$-$$1^{\rm th}$, respectively,  simple root of $SU(N)$, has an 
 asymptotic magnetic field which is  the natural generalization of
(\ref{PSsolution3}) and is given, in the string gauge, by: 
\begin{eqnarray}
\label{BPSBfield}
 B^m\big\vert_\infty &&= {\hat{x}^m \over |x|^2}  \sum_{i=1}^{N-1}   n_i  \left( {\bf \alpha}_i \cdot {\bf H} 
 \right) \cr \cr
 &&={1 \over 2} {\hat{x}^m \over |x|^2}\;  {\rm diag} (n_1 , n_2 - n_1, \ldots , n_j - n_{j-1}, \ldots ,  - n_{N-1})~.
\end{eqnarray}
The asymptotic form of the $SU(N)$ holonomy is as in (\ref{A4infinity}) and the string gauge asymptotics of the gauge field is best described in polar coordinates, with $A_\phi$ its only nonvanishing component:
\be
\label{Afieldinfty}
A_\phi\big\vert_\infty  ={ 1 - \cos \theta  \over 2}  \; {\rm diag} (n_1 , n_2 - n_1, \ldots , n_j - n_{j-1}, \ldots ,  - n_{N-1})~.
\ee
The Kaluza-Klein monopole solution corresponding to the affine root (\ref{affineroot}) will be constructed in Section \ref{kkindex}.

\section{The  index for Dirac operator on $\mathbf{R^3 \times S^1}$}

We define the Callias index of a  Weyl fermion  (with equation of motion $D \psi = 0$ and $D$ defined in (\ref{vdirac})) in the representation ${\cal{R}}$ on $S^1 \times R^3$ 
as in \cite{Callias:1977kg, Weinberg:1979ma}:
\be
\label{indexdef}
I_{\cal{R}}  = \lim\limits_{M^2 \rightarrow 0} {\rm Tr} \;{M^2 \over D^\dagger D + M^2} - {\rm Tr} \;{M^2 \over D D^\dagger + M^2} ~.
\ee
This is the definition most convenient for explicit calculations, despite the fact that  in the locally four dimensional case of interest  an additional regularization will be required, having to do with the need to perform the sum over the Kaluza-Klein tower implicit in (\ref{indexdef}). Nonzero discrete eigenvalues do not contribute to the formal expression (\ref{indexdef})---if $\psi$ is an eigenfunction of $D^\dagger D$ with a nonzero eigenvalue, $D\psi$ is an eigenfunction of $D D^\dagger$ with the same eigenvalue and so their contributions to the trace cancel---hence $I_{\cal{R}}$  counts the number of zero modes of $D$ minus the number of zero modes of $D^\dagger$; the continuous spectrum also does not contribute to (\ref{indexdef}) if all  $\hat{v}_j$ are different, see the discussion   in the appendix of ref.~\cite{Weinberg:1979ma}. The arguments given there continue to hold on $R^3 \times S^1$ and we will not repeat them here---as will become  clear from our results, $I_{\cal{R}}$ of eqn.~(\ref{indexdef}) always yields an integer value for finite action backgrounds on $R^3 \times S^1$. Furthermore, we will show that the index  reduces to the Callias index in the appropriate limit  and experiences discontinuous jumps, which can also be explained physically, upon changing the ratio of the circumference  of $S^1$ ($L$) to the holonomies at infinity ($\hat{v}_j)$ .  

Using the operator $\hat{D}$ from (\ref{vdirac}) and our notation for $\gamma_5$ (\ref{gammas}), we find that:
\be
\label{index2}
I_{\cal{R}}(M^2) = {\rm Tr} \;\gamma_5 {M^2 \over -\hat{D}^2 + M^2 } =  M {\rm Tr} \; \gamma_5 { \hat{D} + M \over -\hat{D}^2 + M^2 }~,
\ee
where the second identity is true because of cyclicity  of trace and $\gamma_5 \hat{D} = - \hat{D} \gamma_5$. Finally we can cancel the $\hat{D} + M$ factor between numerator and denominator and arrive at the expression we will actually use:
\be
\label{index3}
I_{\cal{R}}(M^2) = M {\rm Tr} \; \gamma_5 {1 \over - \hat{D} + M}~.
\ee

In our study, we closely follow the derivation of the Callias index on $R^3$ of ref.~\cite{Weinberg:1979ma}, paying respect to the differences due to the locally four-dimensional nature of spacetime. The main difference---apart from the already mentioned  sum over Kaluza-Klein modes---occurs in the very first step below and has to do with the fact that anomalies occur in a locally four dimensional spacetime. 
 To elucidate, we note that:
\be
\label{qft1}
\langle x \vert {1 \over \hat{D} - M} \vert y \rangle =  \langle \psi(x) \bar\psi(y) \rangle~, 
\ee
where $\langle \ldots \rangle$ denotes an expectation value in a Euclidean quantum field theory of a Dirac fermion $\psi, \bar\psi$ with action $- S = \bar\psi (- \hat{D} + M ) \psi$. For such theories in a locally four dimensional background the following operator identity holds:
\be
\label{qft2}
\partial_\mu J_\mu^5 \equiv \partial_\mu (\bar\psi \gamma_\mu \gamma_5 \psi) = - 2 M \bar\psi \gamma_5 \psi - {T({\cal{R}}) \over 8 \pi^2} G_{\mu\nu}^a \tilde{G}_{\mu\nu}^a~.
\ee
The index (\ref{index3}), via (\ref{qft1}, \ref{qft2}),  can be rewritten as:
\bea
\label{index4}
I_{\cal{R}}(M^2) &=& - M \; \mathrm{Tr} \; \gamma_5 \langle \psi \bar\psi \rangle =  M \int d^3 x \int\limits_{0}^L d y  \; \langle \bar\psi \gamma_5 \psi \rangle \nonumber  \\
&=& -{1 \over 2} \int\limits_{S^2_\infty} d^2 \sigma^k  \int\limits_{0}^L d y \; \langle J_k^5 \rangle - {T({\cal{R}}) \over 16 \pi^2}  \int d^3 x \int\limits_{0}^L d y \; G_{\mu\nu}^a \tilde{G}_{\mu\nu}^a  ~,
\eea
where we used periodicity of the current on $S^1$ to argue that the integral of $\partial_y \langle J_4^5 \rangle$ vanishes.
Eqn.~(\ref{index4})
is our main tool, allowing us to smoothly interpolate the index from $R^3$ to $R^4$ by varying the size of the circle and the appropriate  background.  As a first simple check,  take the limit of an infinite $L$, i.e. $R^4$, where eqn.~(\ref{index4}) becomes:
\be
\label{index5}
I_{\cal{R}}(M^2) =  - {1 \over 2} \int\limits_{S^3_\infty} d^3 \sigma^\mu   \; \langle J_\mu^5 \rangle - {T({\cal{R}}) \over 16 \pi^2}  \int d^4 x   \; G_{\mu\nu}^a \tilde{G}_{\mu\nu}^a ~.
\ee
This is the index theorem\footnote{To avoid (or add) confusion, recall that the index (\ref{indexdef}) for a fundamental Weyl fermion
in an anti-selfdual instanton should be $+1$, as it is $D$, in the notation of Section \ref{notation},  that has a normalizable zero mode.} appropriate for a BPST instanton background (provided that the surface term vanishes:   
that this is so follows from the fact that the surface contribution in (\ref{index5}) could only be due to BPST fermion zero modes, as the nonzero modes vanish exponentially at $S^3_\infty$ and so does their current; the fermion zero modes in an instanton fall off  as a powerlaw  $\psi_0\vert_{x \rightarrow \infty} \sim {\rho\over |x|^3}$, in  nonsingular  gauge  \cite{Vainshtein:1981wh}).

Going back to $R^3 \times S^1$, consider now the integral over $S^2_\infty \times S^1$ in (\ref{index4}). We rewrite the surface term in 
(\ref{index4}) as follows:
\bea
\label{index6}
I_{\cal{R}}^1(M^2) &\equiv& -{1 \over 2} \int\limits_{S^2_\infty} d^2 \sigma^k  \int\limits_{0}^L d y   \; \langle J_k^5 \rangle 
 = -{1 \over 2} \int\limits_{S^2_\infty} d^2 \sigma^k  \int\limits_{0}^L d y \;   {\rm tr} \langle x | \gamma^k \gamma_5 {1 \over - \hat{D} + M}  | x \rangle   \\
&=&-{1 \over 2} \int\limits_{S^2_\infty} d^2 \sigma^k  \int\limits_{0}^L d y \; {\rm tr} \langle x | \left( \gamma^k \gamma_5 \hat{D} {1 \over - \hat{D}^2 + M^2}\right)  | x \rangle~,
\eea
where we performed the operations that led to eqn.~(\ref{index3}) in reverse.
Further, from (\ref{vdirac}), the expressions (\ref{DdaggerD}) for $D^\dagger D$ and $D D^\dagger $, and the explicit form (\ref{gammas}) of the $\gamma$-matrices, we have:
\bea
\label{index7}
I_{\cal{R}}^1(M^2) &=& {1 \over 2}  \int\limits_{S^2_\infty} d^2 \sigma^k \int\limits_{0}^L d y   {\rm tr}  \langle x | \;\sigma^k \sigma^l D_l \left({1 \over -D_\nu^2 + M^2 + 2 \sigma^m B^m} - {1 \over - D_\nu^2 + M^2}  \right)|x  \rangle  \nonumber \\
&-& {1 \over 2}  \int\limits_{S^2_\infty} d^2 \sigma^k \int\limits_{0}^L d y   {\rm tr}  \langle x |\; i \sigma^k D_4 \left({1 \over -D_\nu^2 + M^2 + 2 \sigma^m B^m}+ {1 \over - D_\nu^2 + M^2}   \right)|x  \rangle~,    
\eea
and we recall that (\ref{index7}) is written for an anti self-dual background. 
The final formula for the index which will be used in our further computations is:
\bea
\label{index71}
I_{\cal{R}}(M^2) = I_{\cal{R}}^1(M^2)  - {T({\cal{R}}) \over 16 \pi^2}  \int d^4 x   \; G_{\mu\nu}^a \tilde{G}_{\mu\nu}^a  \equiv I_{\cal{R}}^1 + I_{\cal{R}}^2~,
\eea
with $I_{\cal{R}}^1$ defined in (\ref{index7}) and $I_{\cal{R}}^2$, the topological charge contribution to the index, in (\ref{index71}).
 
 \subsection{The index in a ``static" BPS monopole background and Callias index}

Consider first a 3d BPS ``static" monopole background, independent on the $S^1$ coordinate. 
 Physical intuition tells us that if we consider a small $S^1$, hence weak coupling, we expect the index on $R^3 \times S^1$ to be the same as that on $R^3$ provided  $L v \ll 1$, such that KK modes do not influence physics at scales of order the size of the monopole.  On the other hand, one expects that when $L v \gg 1$, the index can differ from the one on $R^3$. To study how this expectation plays out in detail and under what conditions the index can jump, in the following Sections we successively evaluate the two contributions to the index (\ref{index71}). 

\subsubsection{Surface term contribution}

To evaluate the contribution of the surface term (\ref{index7}), we  note that  at infinity  the dominant terms in the expansion of the operators appearing in $I_R^1(M^2)$ in the  static BPS background are:
\be
\label{infty1}
 -D_\nu^2 + M^2 \simeq - \partial_m^2 + M^2 - D_4^2, ~ {\rm with}~ -i D_4 \rightarrow   {2 \pi n\over L} + A_4 ~,
\ee
where we used the string-gauge asymptotics of $A_4$ (\ref{A4infinity}) and $A_m$ (\ref{Afieldinfty}).
We now expand the surface term contribution, recalling that $B^m \sim r^{-2}$,  observing that only the second term in (\ref{index7}) contributes  after the Pauli matrix traces are taken, and using (\ref{infty1}):
\be
\label{indexstatic3}
I_{\cal{R}}^1(M^2) =   2 \int\limits_0^L d y \int\limits_{S^2_\infty} d^2\sigma^k \; {\rm tr} \; \langle x; y |i D_4 \; {1 \over - \partial_m^2 + M^2 - D_4^2} \; B^k\; {1 \over - \partial_m^2 + M^2 - D_4^2} | x; y\rangle~.
\ee
Next, we substitute the asymptotic form for a ``static" BPS solution, eqn.~(\ref{BPSBfield}), to obtain\footnote{Recall that $n_0 = n_N = 0$ is understood for the static solution.}  using (\ref{infty1}) to replace $i D_4$:
\be
\label{indexstatic4}
I_{\cal{R}}^1(M^2) = - \int\limits_{S^2_\infty} d^2 \sigma^k {\hat{x}_k \over |x|^2} \;  \sum\limits_{p =-\infty}^\infty \sum\limits_{j = 1}^N (\hat{v}_j + {2 \pi p \over L}) (n_j - n_{j-1})  \int { d^3 k \over (2 \pi)^3} {1 \over \left[k^2 + M^2 + (\hat{v}_j+ {2\pi p \over L})^2\right]^2}~.
\ee
After taking the three dimensional momentum and surface integrals
($d^2 \sigma^k \equiv |x|^2   {\hat{x}_k} \;d \Omega_{S^2} $), as well as the $M^2\rightarrow 0$ limit, the surface contribution to the index becomes:
\be
\label{indexstatic5}
I_{\cal{R}}^1(0) = -{1\over 2}\;  \sum\limits_{j = 1}^N (n_j - n_{j-1}) \sum\limits_{p =-\infty}^\infty 
{\hat{v}_j + {2 \pi p \over L} \over |\hat{v}_j + {2 \pi p \over L}|}~.
\ee
The Kaluza-Klein (KK) mode sum in (\ref{indexstatic5}) is a periodic generalization of the sign function, which appears in the Callias index for gauge theories on $R^3$ (upon taking $L\rightarrow 0$ only the $p=0$ term contributes in the sum and so (\ref{indexstatic5}) reproduces the Callias index result, see Appendix \ref{indexformulae}).  Such a generalization is  necessary, since 
on $R^3 \times S^1$ the eigenvalues of the ``Higgs" field $A_4$ are compact and  the index should be a periodic function of the expectation values of $A_4$, with periodicity determined by the representation $\cal{R}$.  

The KK sum (\ref{indexstatic5}) can also be thought of as  a sum of the indices of a KK tower of three-dimensional Dirac operators, each that of a   KK fermion of mass $2 \pi p \over L$. The Callias index theorem shows that for a given Higgs vev  only a finite number of massive operators in the KK tower  have a nonvanishing index (essentially, those with $|m| < {\cal{O}}(|v|)$), thus only  a few terms in the  sum over indices of KK Dirac operators can contribute to the index. While following this logic is a quick way to find our formula for the index for static backgrounds, recall that there is also a non-integer topological charge contribution given by the second term in (\ref{index7}), which should be cancelled by a corresponding non-integer contribution to
(\ref{indexstatic5}) to yield an integer value. 
Thus, to obtain a formula for the index that works for general backgrounds \cite{Gross:1980br}, specified by the holonomy at  infinity, magnetic charge, and topological charge, we must regulate 
the sum over KK modes in (\ref{indexstatic5}). 

For a given $j$,  the KK sum is equal to $\eta_j[0]$, the  spectral asymmetry  of the differential 
operator $h_j = i {d \over d y} + \hat{v}_j$ acting on the space of periodic functions $f(y)=f(y+L)$. The $\eta$-invariant  is defined by analytic continuation from sufficiently large Re$(s) > 0$ of:
\begin{eqnarray}
\label{eta1}
 \eta [ v_j, s] \equiv \eta_j[s] \equiv  \sum_{\lambda \neq 0} {{\rm sign} \lambda \over |\lambda|^{s}},
\end{eqnarray}
where $\lambda$ are the eigenvalues  of $h_j$.\footnote{ An equivalent way to 
 to define the $\eta$-invariant is via its integral representation. Let  $H= i {d \over d y} + A_4$. 
 Then, 
 \be
 \eta[H, s] \equiv  \tr \frac{H}{  (H^2)^{(s+1)/2} } \equiv \frac{1}{\Gamma(\frac{s+1}{2})} \int_{0}^{\infty} \; dt \;  t^{(s-1)/2} \tr [H e^{-H^2 t}]~.
 \ee
 This representation makes sense for  large Re$(s) > 0$  and 
 admits a holomorphic extension to the 
 whole complex plane.   This discussion is completely parallel to much often encountered 
  $\zeta$  function regularization, for which:
  \be
 \zeta[H, s] \equiv  \tr [H^{-s} ] \equiv \frac{1}{\Gamma(s)} \int_{0}^{\infty} \; dt \;  t^{(s-1)} \tr [ e^{-H t}]~.
 \ee 
}
Thus  the surface term contribution to the index is:
\be
\label{indexstatic6}
I_{\cal{R}}^1(0) =- {1 \over 2} \sum\limits_{j=1}^N (n_j - n_{j-1}) \;\eta_j[0]~.
\ee
To calculate $\eta_j[0]$, begin with its definition (\ref{eta1}), rescaling both numerator and denominator by $2 \pi \over L$:
\be
\label{eta11}
\eta_j[s] = \sum\limits_{p=-\infty}^\infty  { {\rm sign} \left({\hat{v}_j L   \over 2 \pi} + p\right) \over | {\hat{v}_j L   \over 2 \pi} + p |^s} = \sum\limits_{p=-\infty}^\infty  { {\rm sign} \left({\hat{a}_j } + p\right) \over | {\hat{a}_j } + p |^s}~. 
\ee
We defined:
\be~~ 
\hat{a}_j \equiv{\hat{v}_j L   \over 2 \pi}  -  \iless{ {\hat{v}_j L   \over 2 \pi}} \subset (0,1)~,
\ee
having noted that since $\eta_j$ is a periodic function of $\hat{a}_j $ of unit period, by relabeling the KK modes, we can take the argument to lie in the fundamental interval $(0,1)$. 
Here 
$\lfloor x \rfloor$ is the floor function: 
\be
{\lfloor x \rfloor}= {\rm max} \{n \in {\mathbb Z} ~|~ n \leq x \} ~,
\ee
which denotes  the largest integer smaller than $x$, and $\hat{x}= x-{\lfloor x \rfloor}$ is the fractional part of $x$.

  It then follows that all terms in the sum (\ref{eta11}) with $p\ge 0$ are positive, while the ones with $p < 0$ are negative, allowing us to write:
\be
\label{eta12}
\eta_j[s] = \sum\limits_{p \ge 0} {1 \over ( \hat{a}_j  + p )^s} - \sum\limits_{p \ge 0} {1 \over (p + 1- \hat{a}_j)^s}  = \zeta(s, \hat{a}_j)  - \zeta(s, 1 - \hat{a}_j)~,
\ee
where $\zeta(s,x)$ is the incomplete zeta-function.
Finally \cite{Gradshteyn}, since $\zeta(0,x) = {1 \over 2} - x$, we find our final expression for $\eta_j[0]$:
\be
\label{eta13}
\eta_j[0] = {1 \over 2} - \hat{a}_j - \left({1\over 2} - (1 - \hat{a}_j)\right) = 1 - 2 \hat{a}_j = 1 - 2  {\hat{v}_j L   \over 2 \pi} + 2 \iless{{\hat{v}_j L   \over 2 \pi} }~.
\ee
For another calculation of the $\eta$-invariant, see Appendix \ref{anothereta}.
 
From (\ref{eta13}), the surface term contribution (\ref{indexstatic5}) to the index for the fundamental representation of $SU(N)$ becomes:
\be
\label{indexstatic81}
I_{fund.}^1(0) =-  \sum\limits_{j = 1}^N (n_j - n_{j-1}) \left({1 \over 2} -   {\hat{v}_j L   \over 2 \pi} + \iless{ {\hat{v}_j L   \over 2 \pi}} \right) ~. \ee

  \subsubsection{Topological charge contribution}

Consider now the second term in (\ref{index71})---the topological charge contribution to the index, which is well-known to be a surface term:
 \be
 \label{indexstatic9}
I_{\cal{R}}^2(0) =  -2 T({\cal R})Q =
-{T({\cal{R}}) \over 16 \pi^2} \int d^3 x \int\limits_0^L d y \; G_{\mu\nu}^a \tilde{G}_{\mu\nu}^a = - {T({\cal R}) \over 16 \pi^2} \int\limits_{0}^L d y \int\limits_{S^2_\infty} d^2 \sigma^m K^m~,
\ee
The topological current   is:
\be
 \label{topocurrent}
 K^\mu = 4 \epsilon^{\mu\nu\lambda\kappa} \tr \left( A_\nu \partial_\lambda A_\kappa + { 2 i \over 3} \; A_\nu A_\lambda A_\kappa \right)~.
 \ee
 In writing the surface integral in (\ref{indexstatic9}), we used the fact that for the static BPS background  $K^\mu$ is  a periodic function of $y$.  
To evaluate (\ref{indexstatic9}) we note that the spatial component of $K^\mu$ can be rewritten as:
\be
K^m = 4 \epsilon^{mij} \tr \left(   A_4 F_{ij} -   A_i \partial_4 A_j  - \partial_i (A_4 A_j)\right)~.
\ee
Now we use $\epsilon^{ijk}F_{jk} = 2 B^i$ and the fact that in the static anti self-dual BPS background (\ref{PSsolution}-\ref{PSsolution3}), assuming $SU(2)$ for now,  
$
8 \tr   A_4 B_m\big\vert_\infty  =  -  8 v {\hat{r}^m \over r^2}  \hat{r}^b \hat{r}^c \tr T^b T^c = - 4 v {\hat{r}^m \over r^2}~.$
Thus, 
 the only contribution to the surface integral (\ref{indexstatic9}) 
 comes from the first   term  in $K^m$, yielding, for $T({\cal{R}})=1/2$:
\be
\label{indexstatic10}
I_{fund., SU(2)}^2(0) =   {1 \over 32 \pi^2} \;  {4 \pi L } \; 4 v =    {L v \over 2 \pi}~.
\ee
This is, of course, the known  result for the negative of the topological charge of an anti self-dual BPS monopole.

To obtain the $SU(N)$ result in the multimonopole background, it is best to transform the surface integral  (\ref{indexstatic9})  to  string gauge and use (\ref{A4infinity}, \ref{BPSBfield}). The singular nature of the static gauge transformation does not change the periodicity of $K_\mu$ used in (\ref{indexstatic9}) and does not affect the surface integral.\footnote{Note that, with $A_\mu^U$=$U A_\mu U^\dagger - i U \partial_\mu U^\dagger$, we have $ K^\mu(A^U)$ = $K^\mu(A)$ + $\epsilon^{\mu\nu\lambda\kappa}\left(  {4 \over 3} \tr (U \partial_\nu U^\dagger \; U \partial_\lambda U^\dagger \; U \partial_\kappa U^\dagger)   + 4 i\; \partial_\nu \tr (A_\lambda  U^\dagger \partial_\kappa U) \right)$ and the singular static gauge transformation does not introduce a shift to $K_m$. \label{variation}}  
Thus, for  an arbitrary representation of $SU(N)$ the topological charge contribution to the index  is: 
\begin{eqnarray}
\label{indexstatic11}
I_{{\cal{R}}}^2(0) &&=  - {T({\cal R}) \over 16 \pi^2} \int\limits_{0}^L d y \int\limits_{S^2_\infty} d^2 \sigma^m  8 \tr [A_4 B_m] \nonumber \\
 && = - 2 T({\cal{R}})   \;\sum\limits_{j=1}^N (n_j - n_{j-1}) {L \hat{v}_j \over  2 \pi }   ~.
\end{eqnarray}

\subsubsection{The final expression for the index}

Combining the two contributions to the index, 
eqns.~(\ref{indexstatic11}) and (\ref{indexstatic81}), gives 
our final formula for the index. Note that neither  the topological charge contribution 
(\ref{indexstatic11}), nor  the surface term (\ref{indexstatic81}) 
is an  integer. However, in the combined result, the non-integer parts 
coming from the two cancel neatly.  With some work, our  expression can also be extracted from the formulae in the Appendix of \cite{Nye:2000eg};  it was derived here in a physicists' manner  by using eqn.~(\ref{qft2}), the axial-current non-conservation which is an exact operator identity valid on any 4-manifold. In this respect, our derivation is a natural generalization of  \cite{Weinberg:1979ma}. 

For the fundamental representation of $SU(N)$, adding (\ref{indexstatic11}) to (\ref{indexstatic81}), the index is:
\begin{eqnarray}
\label{indexstatic12}
I_{fund.}(n_1, n_2, \ldots, n_{N-1}) &&=  - \sum\limits_{j=1}^N  (n_j - n_{j-1})\left( {1 \over 2} +  \iless{{L \hat{v}_j \over   2 \pi }} \right)~ , \nonumber \\&&
=  - \sum\limits_{j=1}^{N-1}  n_j \left(  \iless{{L \hat{v}_j \over   2 \pi }}  -   \iless{{L \hat{v}_{j+1} 
\over   2 \pi }} 
\right)~,
\end{eqnarray}
where in the first line, as usual $n_N=n_0=0$.

It is fairly easy to extract the Callias index theorem from  (\ref{indexstatic12}). Let us restrict
 $ -\pi < L\hat{v}_j <  \pi$ for all $j$. Then,   
 ${1 \over 2} +  \iless{{L \hat{v}_j \over   2 \pi }} =  {1 \over 2} {\rm sign}\;  (\hat{v}_j)$ and    
 (\ref{indexstatic12}) reduces to: 
\be
\label{indexstatic123}
I_{fund.}(n_1, n_2, \ldots, n_{N-1}) = - {1 \over 2} \sum\limits_{j = 1}^N (n_j - n_{j-1})\; {\rm sign}\; (\hat{v}_j)~ = n_{j*}, 
\ee
where ${\hat v}_{j*} <0 <{\hat v}_{j*+1}$ and we used the  the ordering of the holonomies' eigenvalues, eqn.~(\ref{indexstatic123}).  
 In other words the fundamental representation fermion zero mode localizes at the $j^*$$^{\rm th}$ fundamental monopole, the known Callias index result.

\subsection{The index in a ``winding"  BPS-KK monopole background}
\label{kkindex}

Another class of solutions that is crucial for describing the  nonperturbative dynamics  for nonzero $L$ are the Kaluza-Klein monopoles, arising  because of the compact nature of the ``Higgs" field \cite{Lee:1997vp, Kraan:1998pm}; see  \cite{Davies:1999uw} for a semiclassical calculation elucidating their role in supersymmetric gluodynamics.

  Let us recall the construction of the KK monopole   solution corresponding to the ``affine" root
(\ref{affineroot}) of the $SU(N)$ Lie algebra. We will construct the solution in analogy with the simple root monopoles given in 
Section \ref{notation}. To begin, note that we can rewrite the holonomy (\ref{A4infinity}) as follows:
\be
\label{affinevacuum}
A_4 = - \tilde{V} \tau^3 +  {\rm diag}(V, \hat{v}_2 ,\ldots,  ,\ldots , \hat{v}_{N-1}, V)~,
\ee
where now  $V = {1 \over 2} (\hat{v}_N + \hat{v}_1)$ and 
$\tilde{V} =\hat{v}_N - \hat{v}_1$. We  take an $SU(2)$ embedding in $SU(N)$ via $\tau^{1,2, 3}$ 
as:
\begin{eqnarray}
&&(\tau^1)_{ij}= {\textstyle  \frac{1}{2}} ( \delta_{i1} \delta_{jN} + \delta_{iN} \delta_{j1} ),  \qquad
(\tau^2)_{ij} = {\textstyle  \frac{1}{2}} ( -i \delta_{i1} \delta_{jN} + i  \delta_{iN} \delta_{j1} ),  \nonumber  \\
&&(\tau^3)_{ij}= {\textstyle  \frac{1}{2}} ( \delta_{i1} \delta_{j1} -  \delta_{iN} \delta_{jN} ),  \qquad 
i, j=1, \ldots, N.  
\end{eqnarray}
 Clearly, the static self-dual monopole solution is, in complete analogy with the simple-root solutions (\ref{ithembedded}): 
\begin{eqnarray}
\label{affinesoltn}
A_4 &=& \hat{r}^a f(r, \tilde{V}) \tau^a  + {\rm diag}(V, \hat{v}_2 ,\ldots,  \hat{v}_{N-1}, V)~,  \nonumber \\
A_m &=& \epsilon_{m b a} \hat{r}^b g(r, \tilde{V}) \tau^a ~.
\end{eqnarray}

In the class of static solutions (\ref{affinesoltn}) is not a fundamental monopole but can be thought as  a composite of the fundamental solutions based on simple roots. However, in theories with compact Higgs fields it can be used to construct the Kaluza-Klein monopole. 
To begin, note that the non-periodic ``gauge transformation,"\footnote{We use quotation marks as fields related by (\ref{u1}) are not on the same gauge orbit.} defined via our fundamental $SU(2)$ generator $\tau^3$ embedded in $SU(N)$ as described above:
\begin{eqnarray}
\label{u1}
&&U_1(y) = e^{ - i {2 \pi y \over L} \tau^3}  = {\rm diag}(e^{ - i { \pi y \over L} }, 1, \ldots, 1, e^{  i { \pi y \over L} } ) \nonumber \\
&& U_1(y + L) =  
  {\rm diag}( -1, 1, \ldots, 1, - 1 )~ U_1(y) 
\end{eqnarray}
transforms periodic adjoint fields into periodic fields. At the same time, the asymptotic value of $A_4$ is shifted by $U_1(y)$:
\be
\label{u11}
A_4^{U_1} = A_4 + {2 \pi\over L} \tau^3 = - \left(\tilde{V} - {2 \pi \over L}\right) \tau^3 + {\rm diag}(V, \hat{v}_2 ,\ldots  \hat{v}_{N-1},V)~.
\ee
To construct the affine KK monopole, one starts with the static monopole solution (\ref{affinesoltn}) in a vacuum (\ref{affinevacuum}) with $\tilde V$ replaced by $\tilde{V}^\prime = {2 \pi \over L} - \tilde{V}$. Denote by $A_\mu (\tilde{V}^\prime)$, $\mu = (4, m)$,  the just described solution   (\ref{affinesoltn}) in a vacuum given by $\tilde{V}$ $\rightarrow$ $\tilde{V}^\prime$. Then one defines the field configuration:
\be
\label{u2}
A_\mu^{KK}(\tilde{V})= U_2  \left( A_\mu (\tilde{V}^\prime)\right)^{U_1}U_2^\dagger=  U_2 U_1 \left( A_\mu(\tilde{V}^\prime) - i \partial_\mu \right) U_1^\dagger U_2^\dagger~.
\ee
Here $U_2$ is essentially the unit matrix except for its $11$, $NN$, $1N$, and $N1$, elements, explicitly: 
\be
\label{u12}
~~ U_2 \equiv \left( \begin{array}{cccccc} 
0 &  &  &  & &  1 \cr
  & 1&   &&   &    \cr
   &   & 1  & &   &     \cr
  &  &  \ldots & &   &  \cr
  &  &   & & 1&    \cr
-1 &  & &   &  &  0  \cr
 \end{array} \right) ~.
\ee
The point of (\ref{u2}) is that transforming $A_\mu (\tilde{V}^\prime)$ with $U_1$ leads to a twisted (i.e. $y$-dependent) solution in the vacuum  with
asymptotics given by (\ref{affinevacuum}) with $\tilde{V}$ replaced by  $\tilde{V}^\prime - {2 \pi \over L} = - \tilde{V}$. The role of the $U_2$ transformation acting on $A_4$ is to flip 
 the  sign of $\tilde{V}$ and thus generate a solution in the desired vacuum (\ref{affinevacuum}). 
The $A_4$ asymptotics of the KK monopole solution $A_\mu^{KK}$ is thus the desired
(\ref{affinevacuum}), while the $B$-field flips sign at infinity due to the $U_2$ conjugation. Thus the KK monopole solution has   magnetic charge opposite that of the corresponding anti self dual solution---its magnetic charge given by the affine root (\ref{affineroot}) and asymptotics (for  $n_N$ copies of the solution):
\begin{eqnarray}
\label{BPSBfieldKK}
 B^m_{KK}\big\vert_\infty &&= - n_N  {\hat{x}^m \over |x|^2}  \sum_{i=1}^{N}    \left( {\bf \alpha}_i \cdot {\bf H} 
 \right) \cr \cr
 &&={n_N \over 2} {\hat{x}^m \over |x|^2}  {\rm diag} (- 1 , 0, \ldots , 0, \ldots ,  1)~.
\end{eqnarray}

To find the topological charge of the KK monopole, eqn.~(\ref{u2}) plus gauge covariance of the field strength allow us to argue that:
\begin{eqnarray}
\label{u3}
Q&=& {1 \over 32 \pi^2} \int d^3 x d y G^a_{\mu\nu} \tilde{G}^a_{\mu\nu}\left[A^{KK}(\tilde{V})\right] =  {1 \over 32 \pi^2} \int d^3 x dy G^a_{\mu\nu} \tilde{G}^a_{\mu\nu}\left[A^{PS}(\tilde{V}^\prime)\right] \nonumber \\
&=& {L \over 4 \pi^2} \int\limits_{S^2_\infty} d^2 \sigma^m {\rm tr} A_4(\tilde{V}^\prime) B^m_{PS} = - n_N {  \tilde{V}^\prime L \over 2 \pi} =  - n_N \left( 1 - {\tilde{V} L \over 2 \pi}\right) ,
\end{eqnarray}
the calculation in complete analogy with (\ref{indexstatic11}), using the asymptotics of $A_4$, eqn.~(\ref{affinevacuum}) with $\tilde{V}$ $\rightarrow$ $\tilde{V}^\prime$, and  of $B^m_{PS} = - B^m_{KK}$ of (\ref{BPSBfieldKK}). Thus, remembering from eqn.~(\ref{indexstatic9}) that $I_{\cal R}^2 = - 2 T({\cal R}) Q$,  we  obtain that for $n_N$ KK monopoles, the topological charge contribution to the index is:
\be
\label{indexgeneral1}
I_{\cal{R}}^{2,KK}(0) =
 2 T({\cal{R}}) \; n_N \left( 1 -  {\tilde{V} L \over 2 \pi}\right)~.
\ee

The computation of the surface term $I_{\cal{R}}^1(0)$ is also simplified by the fact that the asymptotics of the KK monopole solution at infinity are $x_4$ independent and are, as explained above, the same as those for the PS monopole, except for a switch in the sign of the magnetic field. Thus, despite the fact that in the ``bulk" the solution is twisted around $S^1$, we can still use (\ref{indexstatic3}) to calculate the surface term contribution. Substituting  
eqns.~(\ref{BPSBfieldKK}) and (\ref{A4infinity}) into  (\ref{indexstatic3}),  
we obtain  for the fundamental representation of $SU(N)$, instead of  (\ref{indexstatic81}): 
\be
\label{index1KKfund}
I_{fund.}^{1,KK}(0) =  n_N \left({\tilde{V} L \over 2 \pi} - \iless{{\hat{v}_N L \over 2 \pi}} + \iless{ {\hat{v}_1 L \over 2 \pi}} \right)  ~.
\ee
Combined with (\ref{indexgeneral1}), this gives for  the total index of the KK monopole:
\be
\label{indexKKfund}
I_{fund.}^{ KK}(0) =    n_N \left(1 - \iless{{\hat{v}_N L \over 2 \pi}} + \iless{ {\hat{v}_1 L \over 2 \pi}} \right)  ~.
\ee
In the case where for all $j$ the holonomies obey $|\hat{v}_j| < { \pi \over L}$, 
 taking into account our ordering of the holonomy (\ref{A4infinity}) ($\hat{v}_1 < 0$, $\hat{v}_N>0$), we have, for $n_N=1$, that $I^{KK}_{fund} = 0$. Recall from the discussion around  eqn.~(\ref{indexstatic123})  that in the background of  $n_1, n_2, ..., n_{N-1}$ monopoles corresponding to the $1^{\rm st}$, $2^{\rm nd}$, etc., simple roots
there are $n_{j^*}$ fermionic zero modes, where $j^*$ is the position of the last negative $\hat{v}_j$ from (\ref{A4infinity}). Thus, the combination of a $n_{j^*}=1$ monopole and an $n_N=1$ KK monopole have  a combined number of zero modes equal to that of a four-dimensional BPST (anti) instanton (one for the fundamental of $SU(N)$); the sum of their topological charges also adds to minus one.

At this stage, we can also combine (\ref{indexstatic12}) and  (\ref{indexKKfund}) into a single 
formula: 
\begin{eqnarray} 
I_{fund.}[n_1, n_2, \ldots, n_{N-1}, n_N] &= &
I_{fund.}(n_1, n_2, \ldots, n_{N-1}) +
I_{fund.}^{ KK}(n_N) \nonumber \\
&=&  n_N - \sum\limits_{j=1}^{N}   n_j  \; 
\left( \iless{{L \hat{v}_j \over   2 \pi }}   -    \iless{{L \hat{v}_{j+1} \over   2 \pi }}   \right) ~.
\end{eqnarray}
where $L \hat{v}_{N+1}\equiv L \hat{v}_{1}$.  
 
 \section{${\mathbf {SU(2)}}$ with arbitrary representation fermions}
 
 The calculation of  the index is particularly simple for arbitrary  representations of $SU(2)$. 
  Consider, for example, a Weyl fermion in the  spin-$j$  representation of $SU(2)$  in the   static 
  BPS background. The asymptotic form of the  $A_4$  and magnetic fields are: 
  \be
  A_4\vert_\infty = -v~ (T^3)_j = - v  \; {\rm diag}\;  (j, j-1, \ldots, -j), \qquad 
   B^m\big\vert_\infty  
  = {\hat{x}^m \over |x|^2}\;   (T^3)_j ~,
  \ee
  where we set $n_1=1$ for simplicity. 
The index receives contribution from the surface term  (\ref{indexstatic5}) and topological charge 
(\ref{indexstatic9}).   
 Instead of (\ref{indexstatic5}), we now have:
\be\label{su2anystatic}
I_j^1(0) =  -  \sum\limits_{m=-j}^j m \sum\limits_{p=-\infty}^{\infty} {\rm sign} \left(- vm + {2 \pi p\over L}\right)~,
\ee
where the minus sign in the sign-function is because in our convention the holonomy  at infinity is $A_4 \simeq - v T^3$. 
We perform the KK sum in a way similar to (\ref{eta13}) to obtain:
\be
\label{su2anystatic2}
I_j^1(0) =\sum\limits_{m=-j}^j     - m^2  {v L \over  \pi} - 2m \iless{ -{v m L \over 2 \pi}} ~.
\ee
For the topological charge contribution, we can use the first line of (\ref{indexstatic11}) and following the steps that led to (\ref{indexstatic10}), we obtain:
\be
\label{indexstatic121}
I_{j}^2(0)  = 2 T(j) {L v \over 2 \pi}~.
\ee
Recall that   for the spin-$j$ representation of $SU(2)$, the Casimir is given by 
 $T(j) = \sum\limits_{m=-j}^j  m^2 = {1 \over 3} j (j + 1) (2 j + 1)$. Therefore, summing over the 
 two contributions  (\ref{su2anystatic2}) and (\ref{indexstatic121}) to the index, we find:
\be
\label{indexstaticSU2any}
I_j(0) =  
\sum\limits_{m=-j}^j   2 m \left( - {  m v L \over 2 \pi}  -\iless{- {  m v L \over 2 \pi}} \right) +  2 T(j)\; {L v \over 2 \pi}  = 
-  \sum\limits_{m=-j}^j 2 m \iless{-{m v L \over 2 \pi}}~.
\ee

The relation between the index for the BPS monopole and KK monopole is also especially simple in $SU(2)$, where there are only two kinds of monopoles; in the spin-$j$ representation the index in the KK monopole background can be obtained by using  techniques of the section (\ref{kkindex}), with the result:
\be
\label{su2relation}
I_j^{KK}  = 2 T(j) -I_j
 \ee
where $I_j$ is the index of the $j$-representation in the monopole field and $2 T(j)$ is the number of zero modes in a BPST instanton background. 

Let the number of monopoles and KK monopoles in a given background be, respectively, $n_1$ and $n_2$. The main result of this section is captured in the index  and the topological charge formulae:
\begin{eqnarray}
\label{su2main}
&&I_j [n_1, n_2]  \;  = \;  n_1   I_j + n_2 I_j^{KK} \; =  \; n_2  2 T(j) - (n_1-n_2) 
\sum\limits_{m=-j}^j 2 m \iless{-{m v L \over 2 \pi}} ~,\nonumber \\
&&Q[n_1, n_2] = \;  n_1   Q^{BPS} + n_2 Q^{KK} = - n_2 + (n_2-n_1) \frac{vL}{2 \pi} ~.
 \end{eqnarray}

We consider now as an example the three lowest representations of $SU(2)$. We already  discussed the fundamental representation of $SU(N)$. In the Appendix, we give expressions for other $SU(N)$ representations of interest.

 \subsubsection*{Index for the fundamental ($j=1/2$):}
 
We have, from (\ref{indexstaticSU2any}):
\be
\label{spinhalfsu2}
I_{1/2}(0) =  - \iless{  -{v L \over 4 \pi}} + \iless{{v L \over 4 \pi}}~.
\ee
Begin with  the case  $0 < v < {4 \pi \over L}$, when  we obtain 
  $I_{1/2} = 1$. That this is so can be easily verified by explicitly solving the zero mode equation for the Weyl operator $D$ in the PS background \cite{Jackiw:1975fn}. This is also  the result of the Callias index theorem on $R^3$, as expected on physical grounds when $L$ is small and the scale $v$ of $SU(2)$-breaking is below the KK scale.

Upon increasing $v$, taking  ${4 \pi \over L} < v < {8 \pi \over L}$, we  have   $I_{1/2} = 3$. More generally, eqn.~(\ref{spinhalfsu2}) implies that the index jumps by two every time $v$ crosses another $4 \pi \over L$ threshold. 
This  jump of the index occurs because every time $v$ increases by $4 \pi \over L$, two zero-mode solutions with nonvanishing KK number  become normalizable. This jump of the index can be easily seen explicitly by considering the normalizability of the zero-mode solutions of the  $D(A)\psi = 0$ Weyl equation in the static PS background on $S^1 \times R^3$, along the lines of the Appendix of ref.~\cite{Weinberg:1979zt}.
 
\subsubsection*{Index for the adjoint ($j=1$):}

Now we have from (\ref{indexstaticSU2any}):
\be
I_1(0) = - 2 \iless{-{v L \over 2 \pi}} + 2 \iless{ {v L \over 2 \pi}} ~.
\ee
Begin with $0 < v < {2 \pi \over L}$, where $I_1(0) = 2$, the well-known value in three dimensions. 
As we increase ${2 \pi \over L}< v< {4 \pi \over L}$, we obtain $I_1(0) = 6$. Thus, the index jumps by $4$ every time $v$ crosses a KK threshold. Again, this is because as $v$ passes beyond $2\pi\over L$ every $L=0$ normalizable zero mode acquires two more normalizable KK partners.
  
\subsubsection*{Index for three-index symmetric tensor ($j=3/2$):}
 
Our final example is the three-index symmetric tensor ($j=3/2$ of $SU(2)$). This representation alone is free of a Witten anomaly and gives an example of a chiral four-dimensional theory with interesting non-perturbative dynamics.
The index of the representation is $T(3/2) = 5$.
For this case (\ref{indexstaticSU2any}) implies that the index is:
\be
I_{3/2}(0) =- 3 \iless{- {3 v L \over 4 \pi}} -   \iless{- { v L \over 4 \pi}} + 3\iless{ {3 v L \over 4 \pi} } +  \iless{ { v L \over 4 \pi} } ~.
\ee
For   $0 < v < {4 \pi \over 3 L}$, where $I_{3/2}(0) = 4$, as on $R^3$.
As $v$ increases across the first KK threshold to ${4 \pi \over 3 L} < v < {8 \pi \over 3 L}$, we have  $I_{3/2} = 10$---a jump of the index by $6$. As $v$ crosses the next threshold ${8 \pi \over  3 L} < v < {4 \pi \over  L}$, we similarly find that the index jumps by   6, giving $I_{3/2}(0) = 16$. Similarly to the previous cases, the jumps are interpreted as due to more KK-fermion zero modes becoming normalizable as $v$ increases through each threshold.

 \section{Interpolating from  Callias to APS index }
 It is useful to put together  the results for the index theorem  on $R^3 \times S^1$ and see how it interpolates between the Callias index theorem on $R^3$ and the APS index theorem on $R^4$. This will also provide a crisp  notion of an elementary versus composite topological excitation on 
$R^3 \times S^1$.   In order to study these excitations, it is  useful to recall  
some basic facts about the root system of a Lie algebra  and the distinction between the simple root system and affine root system.  
 
 For a given Lie algebra, we can construct all roots $\Delta$, positive roots  $\Delta^{+}$, and 
  simple positive roots   $\Delta^{0}$, satisfying  $\Delta \supset \Delta^{+} \supset  \Delta^{0} $. 
 For example, all roots  in $\Delta^{+}$ can be written as positive linear combinations  of simple roots which constitute  $\Delta^{0} $:   
 \be
 \Delta^0=\{\alpha_1, \ldots,   \alpha_{N-1} \}~,  
 \ee 
where $\alpha_i $ are $N-1$ linearly independent simple roots. 
The simple root system is 
useful  in  the discussion of the elementary static monopoles, and the discussion of index 
theorems on $R^3$.  

On $R^3 \times S^1$, there is an extra monopole, the KK-monopole, which is on the same footing with the monopoles. The existence of this extra topological excitation is significant in multiple ways. For example, as it will be seen below,  one can only construct the four dimensional 
BPST instanton out of the ``constituent monopoles"   due to the existence of the KK monopole. 
Incorporating the KK-monopole into the set of ``elementary" monopoles also has a simple realization 
in terms of Lie algebra.  There is a unique extended root system (or extended Dynkin diagram) 
 for each  $\Delta^0$, which is obtained by adding the lowest root to the system  $ \Delta^0$:
\be
 \Delta_{\rm aff}^0 =\Delta^0 \cup \{\alpha_N\} \equiv  \{\alpha_1, \ldots,   \alpha_{N-1}, \alpha_N \}
\ee
 
 Let $n_1, \ldots, n_N$ denote the number of elementary monopoles whose charges are 
proportional to $\alpha_1, \ldots, \alpha_N   \in  \Delta_{\rm aff}^0$, respectively. 
The Callias index on $R^3$, for sufficiently small $|\hat{v}_j L|$,  is equal to the index of the Dirac operator on $R^3 \times S^1 $ for elementary monopoles with charges taking values in the simple root system $\Delta_0$, i.e.: 
  \be
 {I}_{R^3}[n_1, \ldots, n_{N-1}, 0] = 
 {I}_{{R^3 \times S^1}} [n_1,  \ldots, n_{N-1}, 0] ~.
   \ee
This is already demonstrated in obtaining (\ref{indexstatic123}) from (\ref{indexstatic12}) by using 
 $|\hat{v}_j L| \leq \pi$. 

We now discuss the relation between the APS index for the BPST instanton and the index theorem on $R^3 \times S^1$. The result is:
  \be
  \label{BPSTmon}
 {I}_{\rm instanton}= 
 {I}_{{R^3 \times S^1}} [1,1,   \ldots, 1,1]=  \sum_{i=1}^N  \; \; {I}_{{R^3 \times S^1}} [0, \ldots, \underbrace{1}_{i^{th}}, \ldots, 0]
   \ee
The proof of this statement necessitates a convenient  rewriting of the 
  index for the ``static"   (\ref{indexstatic12}) and ``winding" (\ref{index1KKfund}) 
  solutions.   The important technical detail to keep in mind is that for  static solutions   
   (\ref{indexstatic12}), we set $n_0=n_N=0$. The index formula for 
   $[n_1, \ldots, n_N]$  monopoles takes the simple form:
\be
\label{indexfund}
I_{fund.}[n_1, \ldots, n_N] = n_N - \sum\limits_{j=1}^{N}   n_j  \; 
\left( \iless{{L \hat{v}_j \over   2 \pi }}   -    \iless{{L \hat{v}_{j+1} \over   2 \pi }}   \right) ~.
\ee

We also need to show that the topological excitation for which  $[n_1, \ldots, n_N]=[1,\ldots, 1]$ 
corresponds to the BPST instanton. It is obvious that the magnetic charge of such an excitation is identically zero, $\sum_{i=1}^{N} \alpha_i=0$.  We also need to show that the topological charge adds up to the one of a BPST instanton. Using formula  (\ref{indexstatic11}) for  static 
BPS monopoles (and setting $n_0=n_N=0$ therein) and  (\ref{u3}) for the KK monopole, 
we obtain the topological charge of the excitation:
\be
Q[n_1, \ldots, n_N]= -n_N + \sum_{i=1}^{N} n_i   \left( \frac{L\hat v_i}{2 \pi} -  
 \frac{L\hat v_{i+1}}{2 \pi} \right) ~.
\ee 
For $[n_1, \ldots, n_N]= k[1, \ldots, 1]$,  the  topological charge is integer valued with no dependence on the specific values of $v_i$. This is indeed the instanton  with winding number 
$k$.   For index theorem aficionados,   eqn.(\ref{BPSTmon}) can also be expressed as:
\be
\rule{0mm}{6mm}
\dim \ker \Dslash_{\rm inst} - \dim \ker \Dslash^{\dagger}_{\rm inst} = 
\sum_{ \alpha_i \in 
\Delta_{\rm aff}^{0}} \left( \dim \ker \Dslash_{\alpha_i} - \dim \ker \Dslash^{\dagger}
_{ \alpha_i} \right) \,.
\ee
It is evident that the index for Dirac operators on $S^1 \times R^3$ has more refined data than 
the familiar APS index theorem for instantons on $R^4$ .

{\bf Remark on some special cases:} In the derivation of the index theorem for the Dirac operator in the background of a monopole, we used the  local axial-current non-conservation (\ref{qft2}), which is an exact operator identity valid on any four-manifold, and a certain boundary Wilson line 
$A_4|_{\infty}$ (\ref{A4infinity}). In fact, the index is only well-defined for invertible $A_4|_{\infty}$. In this case, the corresponding Dirac operator is called a Fredholm operator. 
An eigenvalue of $A_4|_{\infty}$ can always be rotated to zero by turning on an over-all Wilson line as in (\ref{hatv1}), which corresponds to a non-Fredholm operator.
 In those cases, the  index  for the monopole  as well as the $\eta$-invariant are not well-defined. 

What happens physically as the overall $U(1)$ Wilson line is dialed?  In that case, in 
eqn.~(\ref{indexfund}),  we replace $\hat v_j \rightarrow \hat v_j + \frac{1}{\sqrt{2N}}a_0$ following 
eqn.~(\ref{hatv1}). As $a_0$ is dialed smoothly, the fermionic zero mode will jump from the 
monopole it is localized into (say, with charge $\alpha_{j*}$)   to a monopole which is nearest 
neighbor,   $\alpha_{j*\pm1}$, depending on the sign of the $a_0$.  In the mean time, note that 
the index for the BPST instanton 
$ {I}_{\rm instanton}=
 {I}_{{R^3 \times S^1}} [1,1,   \ldots, 1,1]$  in   eqn.~(\ref{BPSTmon})
 should  remain invariant.  
As the normalizable zero mode jumps from  $\alpha_{j*}$ to  $\alpha_{j*\pm1}$, exactly at the value of $a_0$ where one of the eigenvalues becomes zero, a non-normalizable zero mode appears and the exponential decay of the zero mode wave function is replaced by a power law decay of the three dimensional massless fermion propagator.
 
 \section{Remarks on anomalies and induced Chern-Simons terms on $\mathbf{R^3 \times S^1}$}

Consider a chiral four-dimensional gauge theory compactified on $R^3 \times S^1$. In the limit of zero radius, one expects that a generic theory on $R^3$ with complex-representation fermions will violate three dimensional parity. This is because the $R^3$ theory can not be regulated by Pauli-Villars (PV) fields in a  simultaneously parity- and gauge-invariant manner. 

For example, in  the $SU(5)$ theory with left-handed Weyl fermions in the $\mathbf 5$ and $\mathbf{10^*}$, compactified to $R^3$,  four-dimensional Lorentz and gauge invariance would forbid mass terms for the fermions,  but  on $R^3$ real mass terms are allowed. Real mass terms in three dimensions can be thought of as  expectation values of the $A_4$ (Wilson line) components of background $U(1)$ gauge fields gauging global chiral symmetries. These mass terms are gauge invariant but break three dimensional parity. On $R^3$, one can regulate the theory in a gauge invariant manner via real-mass Pauli-Villars fields in the $\mathbf 5$ and $\mathbf{10^*}$. It is a well-known result \cite{Redlich:1983dv} that every PV regulator  gives rise to a Chern-Simons (CS) term, proportional to the index of the representation  ($1/2$ for $\mathbf 5$ and $3/2$ for  $\mathbf{10^*}$) and to the sign of its mass.  Thus, at one loop, the fermion effective action has a parity-violating CS term, whose coefficient is 1 or 2,  depending on the chosen relative sign of the two PV mass terms. This CS term does not give rise to a ``parity anomaly," which would require the addition of a gauge-noninvariant  bare  half-integer coefficient CS term,  since the integer coefficient assures its  invariance under gauge transformations with nontrivial $\pi_3 (G)$ (for a brief reminder of the quantization of the CS coefficient, see the footnote in the beginning of Section \ref{csloop}). However, it gives a topological mass term to the gauge boson. If a  bare  CS term with integer coefficient is added, the CS coefficient becomes a free parameter of the three dimensional ``chiral" gauge theory. When the gauge group is broken to its maximal Abelian subgroup (by an adjoint Higgs field, as in the applications we have in mind) this will give rise to CS terms with quantized coefficients for the various $U(1)$.

\subsection{Loop-induced Chern-Simons terms on $\mathbf{R^3 \times  S^1}$}
\label{csloop}

Now, consider  the same theory on the locally four-dimensional background $R^3 \times S^1$. PV regulators with complex masses are not allowed by gauge invariance, while a real mass due to a Wilson-line expectation value is neither local nor Lorentz invariant. Hence, we are led to reconsider  the calculation of the  CS term, this time on $R^3 \times S^1$. We would like to know whether such a term is generated and what the freedom in the CS coefficient found in the $R^3$ case corresponds to on $R^3 \times S^1$. Our main interest is in the case when the gauge group is broken to its maximal Abelian subgroup by the nontrivial holonomy on $S^1$. Thus, consider the loop-induced CS coefficient $k_{ab}$:
\be
\label{csloop1}
S_{CS} = \int d^3 x \;  {k_{a b}\over 8 \pi}  \; \epsilon_{l i m} \; A_l^a \partial_i \; A_m^b~,
\ee
where   $a$ and $b$ run over the Cartan generators of the gauge group.\footnote{Recall that in the nonabelian case, $S_{CS} = \int d^3 x \;  {k \over 4 \pi}  \epsilon^{l i m} {\rm tr} (  A_l  \partial_i  A_m + {2 i \over 3} A_l A_i A_m)$, where the trace is in the fundamental,  and that $k$ is quantized. To see this, let $U(x)$ denote a gauge rotation for which 
  $\pi_3(G) $ is non-trivial, i.e, $\int  \frac{1}{24 \pi^2} \epsilon^{\nu \lambda \kappa} \tr [U \partial_{\nu} U ^{\dagger}
  \; U \partial_{\lambda} U^{\dagger}\; U \partial_{\nu} U^{\dagger} ] \equiv \int \omega(x) \in {\mathbb Z}$. 
  Under a gauge transformation, the variation of the action is given in footnote (\ref{variation}) and 
  yields $S_{CS}(A^U) = S_{CS}(A) + i (2 \pi k)  \int \omega(x) $, in Euclidean space, showing that  gauge invariance of the partition function demands quantization  of $k$.}
A straightforward loop calculation of $k_{ab}$ in the background holonomy $A_4$  gives:
 \bea
 \label{csloop2}
 k_{ab} &=& \sum\limits_{n=-\infty}^\infty \int {d^3 k \over   \pi^2} \; \tr \; T^a \; {1 \over k^2 + (A_4 + {2 \pi n \over L})^2} \; T^b \; {(A_4 + {2 \pi n \over L}) \over k^2 + (A_4 + {2 \pi n \over L})^2} \nonumber \\
 &=&  \tr \; T^a T^b \sum\limits_{n=-\infty}^\infty \mathrm{sign}\left( A_4 + {2 \pi n \over L} \right)  ,
 \eea
 where a sum over all fermion matter representations is understood in the trace.
To obtain the second equality  we noted that all generators above are in the Cartan and took the momentum integral, leading to a KK sum identical to the one appearing in (\ref{indexstatic5}). Finally, we regulate the sum   via $\zeta$-function as in the calculation of the $\eta$-invariant, and obtain:
\be
\label{csloop3}
k_{ab} =   \tr \; T^a T^b \; \eta[A_4, 0] =
 \tr \; T^a T^b \left(1 - 2  {L  A_4 \over 2 \pi} + 2  \iless{{L   A_4 \over 2 \pi}}  \right)~,
\ee
where the function $\lfloor{...}\rfloor$ is applied to each element of the diagonal matrix $A_4$. To further understand   (\ref{csloop3}), note that  if\footnote{If this condition is not obeyed, the following equations have to  be modified accordingly, as was done in the computation of the index.} $|A_4|< {\pi \over L}$,   we have $ 1 - 2  {L \; A_4 \over 2 \pi} + 2  \iless{{L A_4 \over 2 \pi}}  = - 2 {L A_4 \over 2 \pi}+ {\rm sign} A_4$, and that after inserting this in (\ref{csloop3}) and using $k_{ab}=k_{ba}$, we find:
\be
\label{csloop4}
k_{ab} = - \tr  ( \{T^a, T^b \} A_4 ) {L  \over 2 \pi} + \tr ( T^a T^b {\rm sign} A_4 )~.
\ee

To understand the meaning of the two terms in (\ref{csloop4}), 
we now use the  decomposition of the  sign matrix ${\rm sign}(A_4)$ in each representation $\cal R$ in terms of the unit matrix and Cartan generators:
 \be 
 \label{liedecomposesigna4}
{\rm sign}(A_{4 {\cal R}}) = s^0 1 + \sum_{c=1}^{r} s^c T^c,~~
s^0= \frac{1}{{\rm dim}({\cal R})}\tr_{\cal R} [{\rm sign}(A_4)], ~ ~ s^a= \frac{1}{T({\cal R})}
\tr_{\cal R} [{\rm sign}(A_4) T^a ]~,
\ee 
and  a similar decomposition for the holonomy $A_4$ itself:
\be
\label{liedecomposea4}
 {A_{4} L \over 2 \pi}\bigg\vert_{\cal{R}}= a^0 1 + \sum_{c=1}^{r} a^c T^c,~~
a^0= \frac{L}{2 \pi {\rm dim}({\cal R})}\tr_{\cal R} [ A_4 ], ~~  a^c= \frac{L}{2 \pi T({\cal R})}
\tr_{\cal R} [ A_4 T^c ]~.
\ee
After  inserting these in (\ref{csloop4}),  we find:
\bea
\label{csloop5}
k_{ab} =    \sum_{\cal{R}}  \left[ \tr_{\cal{R}} \left( \{T^a, T^b \} T^c \right) (s^c - a^c ) 
+ T({\cal{R}}) \delta_{ab} (s^0 - a^0) \right]~.
\eea
If $A_4$ is entirely in the Cartan subalgebra of the gauge group, then $a^0 = 0$. Furthermore, if the sign matrix is traceless ($s^0=0$)---which is the case for $SU(2N)$ theories with a center symmetric background---we find that the CS coefficient on $R^3 \times S^1$ is proportional to the coefficient of the gauge anomaly in four dimensions (recall that the anomaly coefficient for a representation ${\cal R}$ is 
 $\tr_{\cal{R}} \left( \{T^a, T^b \} T^c \right)$). In this case, we find that 
 for anomaly-free chiral gauge theories  in four dimensions  there is no loop induced CS term in three dimensions.

It can happen that the sign matrix is not traceless, in which case   the only contribution to the CS term is 
from the second term in (\ref{csloop4}), proportional to tr sign$A_4$.
For example, in anomaly-free $SU(2N+1)$ gauge theories with an almost  center symmetric holonomy,   while the first  term in    (\ref{csloop4}) vanishes, the second term in (\ref{csloop4}) may still be non-zero. In such cases, one can tune a  background Wilson line associated with an axial, non-anomalous $U(1)$ to isolate a point where CS-term vanishes (an example of this kind is $SU(5)$ theory with  $\mathbf 5$ and $\mathbf{10^*}$).

In conclusion, we find that the CS coefficient on $R^3 \times S^1$ receives two contributions. The first is a ``four-dimensional" one and is given by the first term in (\ref{csloop4}). If the only Wilson lines that are turned on correspond to anomaly-free gauge and global symmetries, the contribution of this term vanishes. On the other hand, turning on Wilson lines corresponding to anomalous symmetries leads to a nonvanishing   first term in (\ref{csloop4})---its origin is in the four-dimensional Wess-Zumino term induced when anomalous background fields are included (the reason the calculation in the nontrivial holonomy phase is so simple is that breaking the gauge symmetry and having massive fermions propagate in the loop allowed us to turn the four dimensional Wess-Zumino term into a local three dimensional CS term). The second, ``three-dimensional," contribution \cite{Redlich:1983dv} is given by the second term in  (\ref{csloop4}) and is nonzero only if ${{\rm tr_{\cal{R}}\;  sign} A_4 \over {\rm dim}(\cal{R})}$ generates an anomalous $U(1)$ symmetry in the four-dimensional theory.

\subsection{Excision of topological excitations and remnant Chern-Simons theories}
\label{excision}

In the beginning of this Section, we found that in the three dimensional   reduction of a four dimensional chiral theory, there is freedom to have CS terms with quantized coefficients.  Is there similar freedom in the theory on $R^3 \times S^1$? The answer can again be seen from (\ref{csloop4}). 

In Section \ref{csloop}, we assumed that the only Wilson lines turned on are those corresponding to the Cartan generators of the gauge group. We are free, however, to turn on Wilson lines   of background $U(1)$ fields gauging global  chiral  symmetries in four dimensions. These Wilson lines do not break the  gauge symmetry, but the symmetries they correspond to are usually  anomalous, hence we can use eqn.~(\ref{csloop4}) to infer the CS coefficient induced when they are  turned on. It is clear from (\ref{csloop4}) that, generally, the value of the CS coefficient induced in the nonzero holonomy phase by these ``flavor" Wilson lines would not correspond to quantized values, unlike in three dimensions.
However, recall that turning on Wilson lines for global symmetries is equivalent, by a field redefinition, to imposing non-periodic boundary conditions on the Weyl fermions in $\cal{R}$,\footnote{ For complex representation Dirac fermions, these ``chirally-twisted"  boundary conditions can also be rewritten as
$\Psi(x, y+L)= e^{i \alpha_{\cal {R}} \gamma_5} \Psi(x,y)~.$
 }
\be
\label{chiraltwist}
\psi(x, y+L)_{\cal{R}}= e^{i \alpha_{\cal{R}}} \psi(x,y)_{\cal{R}}~,~~ \alpha_{\cal{R}} = A_4^{{\cal{R}}} L~.
\ee
Consequently,  we find from (\ref{csloop4}), assuming that only a  $U(1)$ Wilson line, $A_4^{_{\cal{R}}}$, is turned on, that:
\be
\label{kab2}
k_{ab}= \delta_{ab} \sum\limits_{\cal{R}} \left(  -   {2 \alpha_{\cal{R}} \;T({\cal{R}}) \over 2 \pi}+ T({\cal{R}}) \; {\rm sign \;\alpha_{\cal{R}}}\right) \equiv \delta_{a b}\;  k(\alpha).
\ee
The induced CS term is, therefore, 
\be
\label{CS}
S_{CS}= \frac{k(\alpha)}{4 \pi} \int_{R^3}  \epsilon^{\nu \lambda \kappa} \; \tr ( A_{\nu}
\partial_{\lambda}A_{\kappa} + \frac{2i}{3} A_{\nu}A_{\lambda}A_{\kappa} ) ~.
\ee
Note that in the case of anomalous-$U(1)$ Wilson lines, the  boundary conditions (\ref{chiraltwist}) would  correspond to a symmetry of the action and measure of  the theory---hence be admissible as boundary conditions---only if the Wilson lines take quantized values,  
\be
\label{quantizedwilson}
2T({\cal R}) \alpha_{\cal{R}} =2 \pi n, ~n  \in {\mathbb Z}~,
\ee
implying that admissible boundary conditions for fermions are quantized (such that the 't Hooft vertex is invariant).  Thus, the coefficients of the induced CS terms in this case also take quantized values.

 Note that  the phase structure of gauge theory---massive versus  perturbatively massless photons---is affected by turning on such discrete Wilson lines. Since the values are quantized, 
 the one-loop potential for the Wilson line  (Casimir energies) should not effect them (discrete Wilson lines  are known to appear in string theory, for example as 
 disconnected components on the moduli space of D-branes \cite{Witten:1997bs}). 
 Moreover, at nonzero $k$, the finite action monopole solutions (or other topological excitations, 
 such as  magnetic bions pertinent to gauge theories on $S^1 \times R^3$)  which would render 
 the gauge fluctuations  massive nonperturbatively do not exist; see, e.g.,~\cite{Seiberg:1996nz}. 
  In this sense, the two types of possible mass terms for gauge fluctuations, parity odd topological CS mass and parity even 
 magnetic monopole or bion induced mass do not mix.  
 
 To summarize, since the chiral anomalous $U(1)$ current is parity odd, the response of gauge theory is to produce a  non-gauge invariant CS term at generic values of the background Wilson line. Only at admissible (discrete set of) boundary conditions for fermions, the induced CS term 
 is gauge invariant and sensible, and a parity odd mass term is generated for the gauge theory. 
 At these points, the finite action topological excitation are excised from the gauge theory. 
If no anomalous $U(1)$ is turned on, the photon is massless to all orders in perturbation theory, and 
a parity even mass term can be induced non-perturbatively via topological excitations 
with zero index, either elementary or composite. 

The notion of the disconnected components of the gauge theory ``moduli space"  may find  interesting 
applications  both  in physics and mathematics. First, we formulate a QCD-like gauge theory on ${\cal M}_3 \times S_1$ where 
 ${\cal M}_3$ is some three-manifold  of arbitrary size, and $S_1$ is small.  Then, we 
   impose  admissible boundary conditions  on the (say) adjoint Weyl fermion\footnote{That this theory is actually SYM plays no role, as one can similarly consider multi-adjoint theories. 
   The discussion can also be generalized to Dirac fermions in complex, two-index 
   representations. For ${\cal R}=   \{\rm AS, S, BF, F \}$ representations, there are respectively, 
   $2T(\cal{R})$ disconnected components (\ref{2TR}), and for   $2T({\cal{R}}) -1$  of them, the long distance physics reduces to pure Chern-Simons theory. In particular, for QCD with one  fundamental fermions, there is no admissible boundary condition for which the infrared physics reduces to CS-theory.} 
 by using a 
   ``chiral twist" (\ref{chiraltwist}) obeying (\ref{quantizedwilson}), by taking $\alpha = {2 \pi n \over 2 N}$, and assuming that $n$ is a positive integer (this is to say that only a ${\mathbb Z}_{2N}$ is a anomaly-free remnant of the 
$U(1)_A$ chiral rotation, and allowed as boundary condition). 
 Integrating out all the heavy KK-modes along the $S_1$ circle induces, among other operators, the CS-term (\ref{CS})  with coefficient given by (\ref{kab2}) and equal to $k(\alpha)= N-n$.  This means that a CS-term does not get induced for strictly periodic and anti-periodic (thermal) boundary conditions. Otherwise, 
  we expect that the long distance dynamics of these disconnected components of the ``moduli space" of  
   QCD-like theories  is described    by  topological CS-theory on   ${\cal M}_3$. This means that
  the theory is gapped, and is in a   topologically ordered Chern-Simons phases.  Up to our knowledge, this is the first derivation of CS-theory and topological phases from QCD-like 
  dynamics. We will pursue this direction in subsequent work.

\acknowledgments
We  thank  M. Shifman  for useful discussions.  This work was supported by the U.S.\ Department of Energy Grant DE-AC02-76SF00515 and by the National Science and Engineering Research Council of Canada (NSERC).

\bigskip

{\flushleft{{\bf Note added:}}} While completing this paper, a new preprint \cite{Ganor:2008hd} appeared, which discusses
the relation between the S-duality and R-symmetry-twisted boundary conditions 
and CS theories in supersymmetric ${\cal{N}} = 4$ SYM theory. In their case, the 3d CS theory arises 
due to mainly S-duality twist and the non-abelian R-symmetry plays a secondary role, essentially 
restoring supersymmetry.  In our mechanism, a  chirally-twisted  boundary conditions associated with 
a discrete anomaly free remnant of axial U(1) symmetry induce the CS term.  
The two mechanism are simply different, but both  yield 3d CS theory in the long distance regime.

   \bigskip 

\appendix

\section{Another calculation of the $\mathbf{\eta}$-invariant}
\label{anothereta}
 Here we give an alternative computation of 
    $\eta_j[0]$ of eqn.~(\ref{eta1}). We now use the  form \cite{AlvarezGaume:1984nf, AlvarezGaume:1985di}:
\be
\label{eta3}
\eta[0] = {1 \over  \pi} \lim_{m \rightarrow \infty} \sum\limits_\lambda {\rm Im } \ln {\lambda + i m \over \lambda - i m} = {1 \over  \pi} \lim_{m \rightarrow \infty}  {\rm Im } \ln \det {h + i m \over h - i m}~,
\ee
which holds because:
$$
\lim_{m \rightarrow \infty}{\lambda + i m \over \lambda - i m} = \lim_{m \rightarrow \infty} e^{i  2 {\rm Arctan}\left({m\over \lambda}\right)} = e^{ i \pi \;{\rm sign} \lambda}~,
$$ and the branch of the logarithm is defined so that $\ln e^{i \phi} = i \phi$ (zero eigenvalues $\lambda$ are assumed to not occur; if they do the formula (\ref{eta3}) is ambiguous and needs to be modified \cite{AlvarezGaume:1985di}). 

Recall from the discussion in paragraph above eqn.(\ref{eta1}) 
that $h$ is the one-dimensional ``massive Dirac operator" $i {d\over d y} + \hat{v}$ whose eigenvalues change sign under  the combined $y \rightarrow - y$, $\hat{v} \rightarrow - \hat{v}$ transformation. Together with (\ref{eta3}) (or (\ref{eta4})) this implies that the spectral asymmetry (\ref{eta3}) flips sign under $\hat{v} \rightarrow - \hat{v}$.
 For our operator, $\lambda = {2 \pi n \over L } + \hat{v}$, so we have:
\bea
\label{eta4}
\eta[0] &=& {1 \over   \pi} \lim_{m \rightarrow \infty} \sum\limits_{n = - \infty}^\infty  {\rm Im } \ln { {2 \pi n \over L} + \hat{v} + i m \over {2 \pi n \over L} + \hat{v} - i m} =  {1 \over  \pi} \lim_{m \rightarrow \infty} \sum\limits_{n = - \infty}^\infty  {\rm Im } \ln { n + a + i m \over n + a - i m}~, \nonumber \\
a &\equiv& {L \hat{v} \over 2 \pi} - \iless{{L \hat{v} \over 2 \pi}} \subset \left(0,1\right)~,
\eea 
where in the first line we  trivially rescaled $m$ and the second line means that $a$ is taken to be in the interval $(0,1)$, which is always possible to achieve by re-labelling the sum over KK modes. We note that the region $(0,1)$ is the fundamental region, as $\eta$ is well-defined and smooth for all points (as opposed to the $(-1/2, +1/2)$ region which includes a singular point $a  = 0$).
The computation of   $\eta[0]$   is simplified by computing the derivative of (\ref{eta4}) wrt $a$. Integration to recover the $a$-dependent part is then trivially done (note that an $a$-independent constant in $\eta[0]$ would be irrelevant, since a $\hat{v}_j$-independent  term in $\eta_j[0]$ does not contribute to the sum in 
(\ref{indexstatic12}); furthermore it is prohibited by the ``parity"-odd nature of $\eta[0]$). The derivative of (\ref{eta4}) wrt $a$ is now given by a convergent sum, which is:
\bea
\label{eta5}
{d \eta[0] \over d a} =  - {1 \over  \pi} \lim_{m \rightarrow \infty} \; 2 m  \; \sum\limits_{n = - \infty}^\infty {1 \over (n+a)^2 + m^2} \equiv - {2\over \pi} \; \lim_{m \rightarrow \infty}  m \; F(1,a,m)~, 
\eea
where the function $F(1,a,m)$ is implicitly defined by the last equality and is computed, e.g., in eqn.~(81) of \cite{Ponton:2001hq}, 
$
F(1,a,m) = {\sqrt{\pi} \over m} ( \sqrt{\pi} + 4 \sum\limits_{p=1}^\infty (\pi p m)^{1 \over 2} \cos(2 \pi p a) K_{1\over 2} (2 \pi p m)) $.
When $m\rightarrow \infty$, only the first term in $F(1,a,m)$  survives,   $\lim_{m \rightarrow \infty}  m F(1,a,m) = \pi$, thus 
$
{d \eta[0] \over d a} = - 2$, 
which determines $\eta_j[0]$ up to an integration constant:
\be\label{eta7}
\eta_j[0] = - 2 a_j + c = - 2 \; \left({L \hat{v}_j \over 2 \pi} - \iless{{L \hat{v}_j \over 2 \pi}} \right) + c .
\ee
This periodic (in $\hat{v}_j$) result can be made ``parity"-odd by taking   $c=1$, giving back (\ref{eta13}).
 
\section{Index for   higher representation  fermions}
\label{indexformulae}

${\bf R^3}$: The Callias index theorem on $R^3$ can be obtained by restricting the sum in (\ref{indexstatic5}) to $p=0$. 
\begin{eqnarray}
{\rm Fund.:}  \qquad I_F(n_1, \ldots, n_{N-1}) &&= - \frac{1}{2}\sum_{i=1}^{N} \;\; 
{\rm sign}( \hat{v}_i) (n_i-n_{i-1})  \nonumber  \\
&&= - \frac{1}{2}\sum_{i=1}^{N-1} \;\; 
n_i \Big[ {\rm sign}( \hat{v}_i ) - {\rm sign}( \hat{v}_{i+1}) \Big]  =n_{j*} ~. \qquad \qquad \qquad \qquad
\label{C1}
\end{eqnarray}
where $ \hat{v}_{j*} <0 <  \hat{v}_{j^*+1}$. 
Since there is no axial anomaly in $d=3$,  there is no other contribution to the index and this is the final result.  A better way to express ( \ref{C1}), which is easily generalizable to arbitrary representation of the gauge group $SU(N)$  is: 
\begin{equation}
I_F(n_1, \ldots, n_{N-1}) = -  \tr[ {\rm sign}(A_4) \cdot \hat B]~.
\label{C2}
\end{equation}
where ${\rm sign}(A_4)$ is the sign matrix and 
$\hat B= \sum_{i=1}^{N-1}   n_i  \left( {\bf \alpha}_i {\bf H} 
 \right) $ is the space independent part of eq.(\ref{BPSBfield}).  For an arbitrary representation  ${\cal R}$, this formula generalizes as: 
 \begin{equation}
I_{\cal R}(n_1, \ldots, n_{N-1}) = -   \tr_{\cal R}[ {\rm sign}(A_4) \cdot \hat B]~.
\label{C3}
\end{equation} 
Our main interest is in  fermionic matter in  two index representations of gauge group, namely, adjoint, antisymmetric (AS), symmetric (S) of $SU(N)$ and 
bi-fundamental (BF) representation of $SU(N) \times SU(N)$. For the adjoint, the index is:
\begin{eqnarray}
\label{C4}
{\rm Adjoint:}  \qquad I_{\rm Adj}(n_1, \ldots, n_{N-1})  &&= 
- \frac{1}{2}\sum_{i, j=1 }^{N} {\rm sign}( \hat{v}_i - \hat{v}_j) \Big[ (n_i-n_{i-1}) - 
 (n_j-n_{j-1})  \Big]  \nonumber  \\
&&= -  \sum_{j=1}^{N-1}  \sum_{i =1 }^{N-1} n_i \Big[ {\rm sign}( \hat{v}_i - \hat{v}_j) - {\rm sign}( \hat{v}_{i+1} - \hat{v}_j) 
   \Big]   
 \nonumber \\  
 &&=  \sum_{j=1}^{N-1} 2 n_j~.
\end{eqnarray}
This means that in the background of each  elementary monopole, there are two fermionic zero modes. For the other two-index representations, the expressions are:
\begin{eqnarray}
&&{\rm BF:}  \qquad I_{ BF}(n_1^1,\ldots, n^1_{N-1},  n_1^2,\ldots, n^2_{N-1})  =- \frac{1}{2}\sum_{i, j=1 }^{N} {\rm sign}( \hat{v}^1_i - \hat{v}^2_j) \Big[ (n^1_i-n^1_{i-1}) - 
 (n^2_j-n^2_{j-1})  \Big]~, \qquad \nonumber \\
&&{\rm AS:}  \qquad I_{ AS}(n_1, \ldots, n_{N-1})  = - \frac{1}{2} \sum_{i > j }^{N} {\rm sign}( \hat{v}_i + \hat{v}_j) \Big[ (n_i-n_{i-1}) + 
 (n_j-n_{j-1})  \Big]~,   \qquad   \nonumber \\
&&{\rm S:}  \qquad I_{ S}(n_1, \ldots, n_{N-1})  = -  \frac{1}{2} \sum_{i  \geq j  }^{N} {\rm sign}( 
\hat{v}_i + \hat{v}_j) \Big[ (n_i-n_{i-1}) +  
 (n_j-n_{j-1})  \Big]~. 
\end{eqnarray}
It is more convenient to express the index for AS/S representations as:
\begin{eqnarray}
I_{ AS/S} &=& - \frac{1}{4} \sum_{i , j=1 }^{N} {\rm sign}( \hat{v}_i + \hat{v}_j) \Big[ (n_i-n_{i-1}) + 
 (n_j-n_{j-1})  \Big] 
 \pm \sum_{i   }^{N} {\rm sign}(2 \hat{v}_i ) (n_i-n_{i-1}) ~. \qquad   \nonumber \\ 
  &=& - \frac{1}{2} \sum_{i , j=1 }^{N-1}  n_i \Big[ {\rm sign}( \hat{v}_i +\hat{v}_j) - 
  {\rm sign}( \hat{v}_{i+1} + \hat{v}_j) 
   \Big]   
 \pm \sum_{i   }^{N-1}  n_i \Big[ {\rm sign}(2 \hat{v}_i ) - {\rm sign}( 2\hat{v}_{i+1}) \Big] 
 \end{eqnarray}

${\bf R^3 \times S^1}:$ This formulae can be straightforwardly generalized to $R^3 \times S^1$ by repeating our derivations for the fundamental. The non-integer contributions to the index from the topological charge cancel the corresponding non-integer part of the $\eta$-invariant, yielding a general expression; further, if the definition of $\hat B$ is extended to include the affine root (\ref{affineroot}), $\hat B= \sum_{i=1}^{N}   n_i  \left( {\bf \alpha}_i {\bf H} 
 \right)$, with $n_1, \ldots ,n_{N-1}$ are the monopole numbers of the background and $n_N$---the KK-monopole number, this equation can be also extended to also include the KK monopole:
\begin{equation}
I_{\cal R}[n_1, \ldots, n_{N}] = 2 T({\cal{R}}) n_N  -   \tr_{\cal R}  \iless{{A_4 L \over 2 \pi}}  \cdot \hat B~.
\label{C5}
\end{equation}
The second term in (\ref{C5}) follows from (\ref{indexstatic3}, \ref{indexstatic81}) by simply extending the definition of $\hat{B}$ to include the affine-root monopole (and by dropping the non-integer terms) while the first term is due to the (negative) integer topological charge contribution to the index of the KK monopole. For reference, 
\be
\label{2TR}
2 T({\cal{R}})= \{1, 2N, N, N+2, N-2\}, \qquad  {\rm for } \; \; {\cal{R}}= \{\rm F, Adj, BF, S, AS \}
\ee
for a single Weyl fermion.  Note that
  \be
 {I}_{ {\cal R}, \rm instanton}= 
 {I}_{ {\cal R}, {R^3 \times S^1}} [1,1,   \ldots, 1,1]=  2 T({\cal{R}}) ~
   \ee
is just the APS index for an BPST instanton.

\end{document}